\documentclass[superscriptaddress, twocolumn, prx, longbibliography]{revtex4-2}
\usepackage[T1]{fontenc}
\usepackage{libertine}
\usepackage{color, amsfonts, amsmath, graphicx, tikz, amssymb, mathtools}
\usepackage{amsthm}
\usepackage[libertine, cmintegrals, bigdelims, vvarbb]{newtxmath}
\usepackage{appendix}
\usepackage{comment}
\usepackage{upgreek}
\definecolor{linkcolor}{rgb}{0,0,0.6}

\definecolor{lgreen} {RGB}{180,210,100}
\definecolor{dblue}  {RGB}{20,66,129}
\definecolor{jblue}  {RGB}{20,50,100}
\definecolor{nblue}  {RGB}{0,120,200}
\definecolor{dgreen} {RGB}{78,138,21}
\definecolor{ngreen} {RGB}{98,158,31}
\definecolor{lred}   {RGB}{220,0,0}
\definecolor{nred}   {RGB}{224,0,0}
\usepackage[pdftex,colorlinks=true,
	pdfstartview = FitV,
	linkcolor    = red,
	citecolor    = blue,
	urlcolor     = blue,	
	hyperindex   = true,
	hyperfigures = false]{hyperref}
\usepackage{cleveref}

\patchcmd{\subsubsection}{\itshape}{\bfseries}{}{}

\usepackage{orcidlink}
\usepackage{dcolumn}
\usepackage{bm}
\usepackage{algorithm}
\usepackage{algpseudocode}
\usepackage{wasysym}
\usepackage{starfont}

\begin{document}

\title{Generalized Finite--time Optimal Control Framework in Stochastic Thermodynamics}
\author{Atul Tanaji Mohite\,\orcidlink{0009-0004-0059-1127}}
\email{atul.mohite@uni-saarland.de}
\affiliation{Department of Theoretical Physics and Center for Biophysics, Saarland University, Saarbrücken, Germany}

\author{Heiko Rieger\,\orcidlink{0000-0003-0205-3678}}
\affiliation{Department of Theoretical Physics and Center for Biophysics, Saarland University, Saarbrücken, Germany}

\begin{abstract}
Optimal processes in stochastic thermodynamics represent a frontier for understanding the control 
and design of non-equilibrium systems, with broad practical applications in biology, chemistry, 
and nanoscale/mesoscale systems. Optimal transport theory and thermodynamic geometry have emerged 
as leading optimal control methodologies, but both rely on slow--driving and close-to-equilibrium 
assumptions. An optimal control framework in stochastic thermodynamics for finite--time driving 
remains elusive. Here, we solve an optimal control problem for driving the control parameters of 
a discrete-state far-from-equilibrium process from an initial to a final value in finite time. 
Optimal driving protocols are derived that minimize the total finite--time dissipation cost of the 
driving process. Our framework reveals that discontinuous endpoint jumps are a generic, 
model-independent physical mechanism that minimizes the optimal driving entropy production --- 
`geometric thermodynamic far-from-equilibrium shortcuts in swift state-to-state transformations' 
--- whose importance is further amplified in far-from-equilibrium systems. The thermodynamic and 
dynamical interpretation of discontinuous endpoint jumps is formulated. An exact mapping between 
the finite--time and slow--driving optimal control formulations is elucidated, advancing the 
state of the art in optimal transport theory and thermodynamic geometry, which has been the 
prevailing paradigm for studying optimal processes in stochastic thermodynamics under slow--driving 
assumptions. Our framework opens up a broad range of applications to the thermodynamically 
efficient control of far-from-equilibrium systems in finite time, and thereby a route towards 
their efficient design principles.
\end{abstract}

\date{\today}

\maketitle
\section{Introduction}\label{sec:introduction}
\subsection{Background and motivation}\label{sec:motivation}
The framework of stochastic thermodynamics (ST) has advanced the thermodynamic understanding of 
microscopic/mesoscopic systems \cite{Shiraishi_2023_book, seifert_2012, sekimoto, Ciliberto_2017, 
ATM_2024_nr_st}, thermodynamically consistent coarse-grained macroscopic systems 
\cite{ATM_2024_nr_st, ATM_2024_nr_cg}, and the trade-off between thermodynamic length, 
dissipation, and fluctuations \cite{atm_2024_var_epr, atm_2025_var_epr_derivation}. In ST, the 
entropy production rate (EPR) quantifies the thermodynamic dissipation cost and constrains the 
dynamics of non-equilibrium systems. For instance, the fluctuation relation (FR) and the 
thermodynamic kinetic uncertainty relation (TKUR) have revealed fundamental thermodynamic laws 
valid beyond the second law of thermodynamics \cite{Shiraishi_2023_book, seifert_2012, sekimoto, 
Ciliberto_2017, ATM_2024_nr_st, Horowitz_2020, atm_2024_var_epr, atm_2025_var_epr_derivation}. 
Recently, a new paradigm of optimal control problems has emerged in ST. Here, the optimal control 
problem is loosely defined as transforming an initial state (control parameter) into a final state 
(control parameter) in finite time, to compute the optimal driving protocols (change of 
state/control parameters) that minimize the thermodynamic dissipation. Three classes of
methodologies have emerged to study optimal control. 

First, optimal processes that minimize finite--time dissipation have been investigated for a 
stochastic particle in a harmonic trap \cite{Schmiedl_2007}. Surprisingly, the finite--time 
optimal protocols were found to exhibit discontinuities at the initial and final times, namely 
the \textit{kinks}. Recently, this mechanism has been understood as a mathematical artefact of 
the imposed boundary condition \cite{Zhong_2024}, inspired by model- and constraint-specific 
methodologies developed for quantum systems \cite{Guery_Odelin_2023} that are valid under 
certain approximations and do not hold for generic cases \footnote{These different methodologies 
are related to different optimal control formulations of classical stochastic systems discussed 
subsequently, but they are not necessarily the same. Fundamental differences between them can be 
drastic, despite their case-specific applicability. First, their specific aims differ 
significantly depending on what they are designed for, mainly the dynamical aspects; see 
Ref.~\cite{Guery_Odelin_2023}. Second, the non-equilibrium notion of entropy production rate 
itself is not necessarily well defined for quantum thermodynamic systems, but it is of paramount 
importance for classical stochastic thermodynamic systems. As a result, these quantum 
methodologies strictly focus on the optimization of dynamical aspects, such as constraints on 
the dynamical evolution of the quantum state or operators, or on their eigenstates.

$\quad$ These methods justify the importance of dynamical aspects over thermodynamic aspects for 
finite--time optimal processes, arguing that dynamics become more significant than thermodynamics. 
However, by construction, the most fundamental principles of statistical physics imply that 
thermodynamics constrains the dynamics, and not the other way around. Therefore, even the 
dynamics of finite--time optimal processes \textit{must} be constrained by thermodynamics. We 
rigorously prove this within our framework.}. However, these works fail to capture the physical 
and thermodynamic origin of the \textit{kinks} \cite{Schmiedl_2007, Zhong_2024}. The 
universality of \textit{kinks} has been observed in some computationally solvable models 
\cite{Gomez-marin_2008, Then_2008, Bonaca_2018, Engel_2023}. Despite a few model-specific 
studies, analytically computable solutions for other models are lacking, which has created a 
void in the theoretical physical understanding/framework to attribute this phenomenon.

Second, thermodynamic length defines the distance in the control parameter space of a model and 
connects it to the thermodynamic dissipation cost \cite{Rao_1945,Weinhold_1975, Ruppeiner_1995, 
Salamon_1985, Schlogl_1985, Brody_1995, Crooks_2007}. This Riemannian geometric structure in 
the control parameter space has been exploited to compute optimal driving protocols and formulate 
the framework of thermodynamic geometry, which is valid in the slow--driving limit. Within this framework, a geodesic is a minimum-distance path between the initial and final control parameters 
and is equivalently the optimal driving protocol \cite{Crooks_2007, Sekimoto_1997_slow_driving_optimal_control, 
Sivak_2012}. Thermodynamic geometry has advantages due to its practical applicability. In 
particular, metric tensors are numerically and experimentally computed for sophisticated models 
and systems \cite{Crooks_2007, Sekimoto_1997_slow_driving_optimal_control, Sivak_2012, 
Mandal_2016, Sivak_2016, Zulkowski_2012, Zulkowski_2013, Zulkowski_2014, Zulkowski_2015, 
Rotskoff_2017, Loutchko_2022_prr_riemanian, Chen_2021_pre, Li_2022, Watanabe_2022, Ito_2020, 
Li_2022}. Thermodynamic geometry has therefore been rigorously studied in comparison to the 
first methodology, owing to its practical applicability. However, thermodynamic geometry has two 
major drawbacks. First, it relies on the slow--driving approximation, which makes it suboptimal 
for finite--time optimal driving processes where the driving time is small and the slow--driving 
assumption is inherently violated; see Ref.~\cite{Ma_2020} for an experimentally verified 
violation. Second, it lacks \textit{kinks} in the optimal driving protocols, rendering it 
inconsistent with exact analytical solutions in Refs.~\cite{Schmiedl_2007, Zhong_2024, 
Gomez-marin_2008, Engel_2023, Then_2008, Bonaca_2018}. Thermodynamic geometry is a sub-case of 
the general mathematical framework known as information geometry; however, even information 
geometry does not address these drawbacks \cite{Rao_1945,Amari_2016_book}.

Recently, a third optimal control methodology has emerged in ST, which uses a mathematical 
framework: optimal transport theory \cite{Jordan_1998, Benamou_2000, villani, omtp-book}. The 
insights from optimal transport theory have been incorporated into ST \cite{Aurell_2011, 
Aurell_2012, Aurell_2012-landauer, Chen_2019_stochastic_control, Zhang_2020, Zhang_2020_engine, 
Dechant_2019, Nakazato_2021_omtp_st, Fu_2021_wasserstein-engine, Taghvaei_2021, Van_vu_2023, 
Ito_2024_omtp_st, Sabbagh_2024, Klinger_2025}. The mapping between ST and optimal transport 
theory relies on the equivalence between the EPR in ST and the Wasserstein distance in optimal 
transport theory, an information-theoretic distance measure between probability distributions. 
The optimal transport map corresponds to optimally transforming an initial probability 
distribution into a final probability distribution; equivalently, the optimal driving protocol 
is obtained here. Hence, optimal control in optimal transport theory assumes `full' control of 
the probability distribution defined in the state space; in contrast, thermodynamic geometry 
assumes a parametric control defined in the control parameter space. Due to its inherent 
formulation as an optimization problem, optimal transport theory has a broad range of 
applications in statistical machine learning and computer science \cite{Kolouri_2017}, including 
computer vision, linguistics, signal processing, and image representation. It is also used to 
solve nonlinear differential equations \cite{omtp-book}, which are of paramount importance in 
physics \cite{Jordan_1998}, biology, economics, and fluid mechanics \cite{Benamou_2000}.

Despite its multitude of successful applications, optimal transport theory has three major 
drawbacks. First, despite the numerical applicability of optimal transport theory, exact 
analytical solutions are not available except for Gaussian systems, a statistical approximation 
that does not necessarily hold for finite-size systems in ST prone to non-Gaussian fluctuations 
\cite{atm_2024_var_epr, atm_2025_var_epr_derivation}. Second, the quadratic dependence of EP 
on the transition affinity is an assumption that holds for close-to-equilibrium (cEQ) systems 
\cite{atm_2024_var_epr, atm_2025_var_epr_derivation, ATM_2024_nr_st, ATM_2024_nr_cg} and leads 
to a massive underestimation of EP for far-from-equilibrium (fEQ) systems 
\cite{atm_2024_var_epr, atm_2025_var_epr_derivation}. Third, when the driving time to reach 
from the initial to the final state is finite, the driving time is a resource to be optimized 
over; optimal transport theory does not take this constraint into account, resulting in 
suboptimal performance for finite--time optimal processes, since optimal transport theory is 
built upon the slow--driving assumption. Therefore, the results obtained using optimal transport 
theory are inconsistent with the model-specific exact analytical results of 
Refs.~\cite{Schmiedl_2007, Gomez-marin_2008}; the \textit{kinks} are absent even for the 
simplest models --- see Ref.~\cite{Oikawa_2025} for an experimental comparison.

The total EP has three linearly independent contributions, namely driving work, excess EP, and 
housekeeping EP \cite{ATM_2024_nr_st, atm_2025_var_epr_derivation}, which physically correspond 
to dissipation due to the total free energy, relaxation towards the Boltzmann distribution, and 
sustaining nonconservative forces, respectively. However, in the works discussed so far, the 
focus of optimization has been on driving work or excess EPR, while the optimization of the 
housekeeping EPR --- which quantifies the thermodynamic cost of sustaining non-equilibrium 
currents --- has been entirely absent. In addition, the focus has been on continuous-state 
systems rather than discrete-state systems, where the non-quadratic dependence of the EPR on 
the driving affinity renders close-to-equilibrium optimal control methods physically less 
relevant, as they assume a quadratic dependence of the EPR on the driving affinity and neglect 
the housekeeping EPR \cite{atm_2024_var_epr, atm_2025_var_epr_derivation}. Despite attempts to 
understand finite--time optimal processes consistently within ST, inconsistencies and discrepancies 
persist, and a coherent, unified framework for finite--time optimal control in ST is lacking. 
\subsection{Setup}\label{sec:setup}
\subsubsection{Thermodynamically consistent discrete--state systems and graphs} 
We model thermodynamically consistent discrete-state systems using a Markov jump process (MJP) 
or a chemical reaction network (CRN), represented by a graph \cite{schnakenberg_1976}.
{Throughout this work, the notion of thermodynamic consistency refers to the local detailed balance condition (LDB), which in ST offers an equivalence between the dynamical and thermodynamic properties of the system.} 
$\rho_i$ denotes the probability/density of the state, which is a node of the graph. The set 
of all discrete states is denoted by $\{i\}$. $\gamma^{\rightleftharpoons}$ denotes the set of 
forward and backward transitions between states $\rho_i$ and $\rho_j$, with $J_\gamma$ and 
$A_\gamma$ denoting the current and affinity for the transition $\gamma^{\rightleftharpoons}$ 
between states. $\{\gamma^{\rightleftharpoons}\}$ denotes the set of all bidirectional 
transitions of the graph. The transitions satisfy the LDB, 
$A_{\gamma} = \ln(j_{\gamma}/j_{-\gamma}) = F_\gamma - \Delta_{\gamma} E + \Delta_{\gamma} 
S^{state}$, where $j_{\gamma}$ and $j_{-\gamma}$ are the forward and backward transition 
currents satisfying $J_\gamma = j_\gamma - j_{-\gamma}$.

The transition affinity $A_{\gamma}$ is decomposed into an external non-conservative driving 
$F_{\gamma}$, a change in the equilibrium energy functional $\Delta_{\gamma}E$, and a change 
in the state entropy $S_{i}^{state} = -\ln(\rho_i)$ \cite{seifert_2012}.
{We measure $E$ and $F_\gamma$ in units of the inverse temperature $\beta$, and set $k_\text{b} = 1$. $-\Delta_{\gamma}E$ is equivalently defined as the difference between the chemical potential of the reactant state ($\rho_{r_{\gamma}} \in \{i\}$) and the product state ($\rho_{p_{\gamma}} \in \{i\}$); $-\Delta_{\gamma}E = \mu_{r_\gamma} - \mu_{p_\gamma}$. $F_{\gamma}$ is a property of the transition; therefore, coupling to different thermodynamic reservoirs is modelled by using different transition channels between the same states. Similar modifications also apply to the inverse temperature $\beta$ when different reservoirs are at different temperatures. Throughout this work, we use the vector ($\vec{*}$) and equivalent set ($\{*\}$) representations interchangeably, depending on the context. $|*|$ denotes the dimension of a set or vector. $\rho_i^E = e^{-E_i + \psi_E}$ is the equilibrium Boltzmann distribution, with the corresponding equilibrium free energy $\psi_E = -\ln\!\left(\sum_{\{i\}} e^{-E_i}\right)$ defined for the Boltzmann distribution. The energy functional $E(\{\lambda_E\})$ (and therefore $\psi_E(\{\lambda_E\})$ also) is fully controlled by the set of linearly independent control parameters $\{\lambda_E\}$ of $E$, with dimension $|\{\lambda_E\}|$. The dimension of the state space is equal to $|\{i\}|$ and the dimension of the transition space is equal to $|\{\gamma^{\rightleftharpoons}\}|$.}

The symmetric component of the transition currents is called the traffic and is defined as 
$T_\gamma = j_\gamma + j_{-\gamma}$. The transition mobility is defined as 
$D_\gamma = \sqrt{j_\gamma j_{-\gamma}}$; it is time-symmetric, quantifying the amplitude (or 
equivalently the inverse timescale) of the transition currents. This allows us to define 
hyperbolic relations between different bases $\{A_\gamma, D_\gamma\} \to \{J_\gamma, T_\gamma\}$: 
$J_\gamma = 2D_\gamma \sinh(A_\gamma/2)$ and $T_\gamma = 2D_\gamma \cosh(A_\gamma/2)$. The 
bases $\{J_\gamma, T_\gamma\}$ and $\{A_\gamma, D_\gamma\}$ formulate the thermodynamic 
inference and full control descriptions, respectively. This nomenclature is attributed to the 
affinities $\{A_\gamma\}$ and the mobilities $\{D_\gamma\}$ being the controllable physical 
parameters, which physically correspond to controlling the current asymmetry and amplitude, 
respectively \cite{ATM_2024_nr_st}.

We introduce a shorthand notation for the state--space column vector $\vec{\rho} = (\ldots, \rho_i, \ldots)^T$ 
and the current-space vector $\vec{J} = (\ldots, J_\gamma, \ldots)^T$. The continuity equation 
for the transport of probabilities/densities is
\begin{equation}\label{eq:continuity_equation}
\begin{split}
    \partial_t \vec{\rho} = \mathbb{S} \vec{J}.
\end{split}    
\end{equation}
The stoichiometry matrix $\mathbb{S}$ encodes the topology of the transition space 
$\{\gamma^{\rightleftharpoons}\}$ and contracts the transition currents $\{J_\gamma\}$ to the 
state space $\{\rho_i\}$. The entries $\mathbb{S}_{i\gamma}$ of $\mathbb{S}$ are $1$ (for 
product) or $-1$ (for substrate) if state $\rho_i$ is part of a transition 
$\gamma^{\rightleftharpoons}$, with the sign convention decided by the direction of the 
transition currents; otherwise, $\mathbb{S}_{i\gamma} = 0$.
{
To demonstrate the notation, consider a prototypical example of two protein conformation states, 
denoted by `1' and `2', that can undergo transitions through two different channels, coupled to 
an `equilibrium' and an `enzymatic' (non-equilibrium) reservoir, respectively. The state space 
and transition space are thus $\{i\} = \{1, 2\}$ and $\{\gamma^{\rightleftharpoons}\} = \{eq, enz\}$. 
The transition affinities for the `equilibrium' and `enzymatic' transitions are
$A_{eq} = \mu_1 - \mu_2 + \ln(\rho_1) - \ln(\rho_2)$ and 
$A_{enz} = F_{enz} + \mu_1 - \mu_2 + \ln(\rho_1) - \ln(\rho_2)$, respectively, 
which equivalently quantify the thermodynamic cost of a microscopic transition supported by 
the `equilibrium' and `enzymatic' reservoirs.
}
\subsubsection{Thermodynamic dissipation of graphs}
The mean EPR for the graph satisfies the bilinear form $\langle\dot{\Sigma}\rangle = 
\sum_{\{\gamma^{\rightleftharpoons}\}} J_\gamma A_\gamma = \sum_{\{\gamma^{\rightleftharpoons}\}} \Sigma_\gamma$, which physically corresponds to the 
thermodynamic dissipation being equal to the driving force (affinity) multiplied by the current 
it generates. Furthermore, the mean EPR $\langle\dot{\Sigma}\rangle$ for the graph is 
decomposed into three linearly independent orthogonal contributions 
\cite{ATM_2024_nr_st, atm_2025_var_epr_derivation},
\begin{equation}\label{eq:mean_EPR}
\begin{split}
    -\dot{\psi}_E & = -  \dot{\lambda}_E \: \partial_{\lambda_E} \psi_E,
    \\
    \langle \dot{\Sigma}_E^{ex} \rangle & = -\sum_{\{i\}} d_t \rho_i \ln{\left( \frac{\rho_i}{\rho_i^E} \right)} = -d_t D_E^{KL},
    \\
    \langle \dot{\Sigma}^{hk} \rangle & = \sum_{ \{ \gamma^\rightleftharpoons \} } T_\gamma^\perp F_\gamma \sinh{ \left( \frac{F_\gamma}{2} \right) } = \sum_{ \{ \gamma^\rightleftharpoons \} } \langle \dot{\Sigma}_\gamma^{hk} \rangle.
\end{split}
\end{equation}
These are the driving of free energy (quasi-static work rate), excess EPR, and housekeeping EPR, 
respectively, corresponding to the driving of the energy functional $E$ through a set of 
external control parameters $\{\lambda_E\}$, the relaxation towards the Boltzmann distribution 
dictated by $E$, and the sustained dissipative transition currents due to non-conservative 
forces $\{F_\gamma\}$, respectively\footnote{This threefold orthogonal decomposition of the mean 
dissipation was highlighted in Ref.~\cite{atm_2025_var_epr_derivation} and generalized to a 
fourfold orthogonal decomposition in Ref.~\cite{ATM_2024_nr_st} for `actio=reactio' 
symmetry-breaking non-reciprocal systems.}.
{
Note that the choice of energy functional $E$ is flexible and corresponds to a `gauge fixing' 
dictated by experimental or physical constraints. Therefore, it does not necessarily have to be 
an equilibrium energy functional; for instance, when a non-equilibrium analogue of energy is used 
for a non-equilibrium steady state (NESS), $E = ss$, such as a Lyapunov functional or a rate 
functional that generates the NESS distribution, $\psi_{ss}$ is the corresponding 
non-equilibrium free energy analogue for the NESS. This geometric definition therefore broadly incorporates the exponential family of state--space Boltzmann distributions.
}

$-\dot{\psi}_E$ and $\dot{\Sigma}_E^{ex}$ are boundary terms in the control parameter space 
and the state space, respectively \cite{ATM_2024_nr_st, atm_2025_var_epr_derivation}. 
$-\dot{\psi}_E$ is integrated over time to obtain the free-energy difference between the 
initial and final control parameters, $ \Delta_{\lambda_E} \psi_E= \psi_E(\vec{\lambda}_E^{inl}) - \psi_E(\vec{\lambda}_E^{fnl})$. 
Similarly, the Kullback--Leibler (KL) divergence ($D_E^{KL}$) quantifies the statistical 
distance of the instantaneous state--space distribution from the equilibrium Boltzmann 
distribution, $\rho_i^E = e^{-E_i + \psi_E}$. Thus, a time-integrated excess EPR yields an 
excess EP, defined as the difference between the initial and final state--space distributions 
with respect to the initial and final control parameters,
$-\Sigma_E^{ex} = \Delta_{\lambda_E} D_E^{KL} =  
D_{E}^{KL}(\vec{\lambda}_E^{fnl}, t=\tau^{obs}) - D_{E}^{KL}(\vec{\lambda}_E^{inl}, t=0)$, with 
$D_{E}^{KL}( \vec{\lambda}_E^{inl}, t=0) = \sum_{\{\rho_i\}} \rho_i^{inl} 
\ln[ \rho_i^{inl}/\rho_i^{E(\vec{\lambda}_E^{inl})} ]$ and 
$D_{E}^{KL}( \vec{\lambda}_E^{fnl}, t=\tau^{obs}) = \sum_{\{\rho_i\}} \rho_i^{fnl} 
\ln[\rho_i^{fnl}/\rho_i^{E(\vec{\lambda}_E^{fnl})}]$.

The mean housekeeping EPR $\langle\dot{\Sigma}^{hk}\rangle = \sum_{ \{ \gamma^\rightleftharpoons \} } \langle \dot{\Sigma}_\gamma^{hk} \rangle$ quantifies NESS dissipation, $\langle \dot{\Sigma}_\gamma^{hk} \rangle$ is a non-quadratic function of the 
non-conservative driving force $F_\gamma$, multiplied by the equilibrium traffic $T_\gamma^\perp$ 
(the component orthogonal to the external driving) \cite{ATM_2024_nr_st, atm_2025_var_epr_derivation}. 
$T_\gamma^\perp$ quantifies the scaled equilibrium diffusion constant defined for 
$\gamma^{\rightleftharpoons}$ due to the equilibrium thermodynamic activity of states $\rho_i$ 
and $\rho_j$. For example, for MJP/ideal CRN, $T_\gamma^\perp = \rho_i + \rho_j$, and for 
interacting CRN, $T_\gamma^\perp = \rho_i e^{\mu_i^{int}} + \rho_j e^{\mu_j^{int}}$, where 
$\mu_i^{int}$ is the chemical potential of $\rho_i$ attributed to the non-ideal interactions 
of $\rho_i$ \cite{ATM_2024_nr_st, ATM_2024_nr_cg}.
{
The set $\{A_\alpha\} = \{\{\lambda_E\}, \{-\ln(\rho_i/\rho_i^E)\}, \{F_\gamma\}\}$ defines all 
linearly independent (orthogonal) control parameter affinities of the graph, whose dimension is 
equal to $|\{\lambda_E\}| + |\{i\}| + |\{\gamma^{\rightleftharpoons}\}|$. $\{A_\alpha\}$ accounts 
for the orthogonal decomposition of the EPR, in contrast to $\{A_\gamma\}$. We use the subscript 
$\alpha$ for the controllable orthogonal transition affinities $A_\alpha$, distinguishing them 
from the microscopic transition affinities $A_\gamma$ of the model. Importantly, integrating 
\cref{eq:mean_EPR} over the observation time $\tau^{obs}$ implies, $-\Delta\psi_E \propto 
O(1)$, $\langle\Sigma_E^{ex}\rangle = -\Delta D_E^{KL} \propto O(1)$, and 
$\langle\Sigma_E^{hk}\rangle \propto O(\tau^{obs})$, revealing the short-time and long-time 
symmetries of the mean EPR $\langle\dot{\Sigma}\rangle$: regimes characterized by the dominance 
of $\Delta\psi_E$ and $\langle\Sigma_E^{ex}\rangle$ (short time) and $\langle\Sigma^{hk}\rangle$ 
(long time) \cite{atm_2025_var_epr_derivation}.
}

{
We briefly highlight how the applicability of the setup extends to continuous state--space systems with overdamped Langevin type dynamics, 
which are obtained by coarse-graining the underlying microscopic discrete-state systems 
\cite{ATM_2024_nr_cg}. Continuous state--space systems satisfy analogous dynamics [a Fokker--Planck 
equation for state--space probability transport, instead of \cref{eq:continuity_equation}] and 
thermodynamic [the threefold decomposition of the mean EPR in \cref{eq:mean_EPR}] structures, 
where the free energy $\psi_E$ and excess EP $\langle{\Sigma}^{ex}\rangle$ are defined (computed) 
over the continuous space measure. Due to transition--space coarse-graining, 
$\langle\dot{\Sigma}^{hk}\rangle$ due to a non-conservative drift term for continuous state--space systems is defined in state space, 
and its familiar quadratic dependence is recovered in the small-$F_i$ approximation [$F_\gamma$ 
in \cref{eq:mean_EPR}; $\sinh(x) \approx x$ for small $x$]. However, the exact non-quadratic 
dependence is necessary to preserve the microscopic dissipation of continuous state--space systems 
beyond the length--scale of the underlying microscopic discrete-state systems 
\cite{ATM_2024_nr_st, ATM_2024_nr_cg}. Due to sophistication associated with underdamped Langevin dynamics, their study is out of the scope of this work.
}
\subsubsection{Minimum action principle}
Due to their favourable computational aspects, variational formulations of physical processes 
have been employed to study the dynamics of non-equilibrium systems --- for instance, phase 
transitions, first-passage times, and metastability in stochastic systems 
\cite{Weinan_2004, Heymann_2008, Vanden-Eijnden_2008, Grafke_2017, Grafke_2019, Gagrani_2023, 
Zakine_2023, Smith_2024} --- as well as efficient numerical optimization problems in machine 
learning \cite{Cherukuri_2017, Lin_2020, Yan_2022}, and have recently been explored in ST 
\cite{kim_2020, Otsubo_2020, Manikandan_2020, Otsubo_2022, Horiguchi_2024, Tottori_2024}. 
Recently, an exact variational formulation for fEQ discrete-state processes has been developed, 
namely the `minimum action principle' (MinAP) \cite{atm_2024_var_epr, atm_2025_var_epr_derivation}, 
an exact canonical ensemble analogue of non-equilibrium discrete-state systems that unifies the 
FR and the non-quadratic formulation of the TKUR within a single framework 
\cite{atm_2025_var_epr_derivation}. 

The transition probability measure for discrete-state processes is equal to the exponential of 
an action $\mathcal{S}\left[\{A_{\gamma}, D_\gamma\}\right]$ 
\cite{atm_2024_var_epr, atm_2025_var_epr_derivation, ATM_2024_nr_cg},
\begin{equation}\label{eq:transition_probability_measure}
\begin{split}
    \mathcal{P}\left[\{A_{\gamma}, D_\gamma\}\right]
    = e^{-\mathcal{S}\left[\{A_{\gamma}, D_\gamma\}\right]},
\end{split}    
\end{equation}
where the action $\mathcal{S}\left[\{A_{\gamma}, D_\gamma\}\right] = \int_0^\tau \mathcal{L}^*[\{A_{\gamma}, D_{\gamma}\}] \, dt$ 
is a time integral of the effective transition Lagrangian, which equivalently quantifies the 
effective mean EPR,
\begin{equation}\label{eq:lagrangian_full_contol_description}
\begin{split}
    \langle\dot{\Sigma}\rangle = \mathcal{L}^*[\{A_{\gamma}, D_{\gamma}\}]   
    & = \sum_{\{\gamma^{\rightleftharpoons}\}} 2D_\gamma A_\gamma \sinh\!\left(\frac{A_\gamma}{2}\right),
\end{split}    
\end{equation}
represented here in the `full' control description using the transition mobility $D_\gamma$ and 
the transition affinity $A_\gamma$. \Cref{eq:transition_probability_measure,eq:lagrangian_full_contol_description} 
are key results of Refs.~\cite{atm_2024_var_epr, atm_2025_var_epr_derivation} and formulate a 
variational principle for discrete-state processes. In particular, the analytical solution for 
the dynamics of discrete-state processes is obtained by analytically or numerically solving the 
variational problem for $\mathcal{L}^*[\{A_{\gamma}, D_{\gamma}\}]$, namely the `minimum action 
principle' (MinAP).
{
Since the definition of $A_\gamma$ explicitly requires the `forward' and `backward' transitions, 
the transition probability measure in \cref{eq:transition_probability_measure} is defined with 
respect to the conjugate dynamics that is `microscopically time-reversed' relative to the original 
dynamics \cite{atm_2025_var_epr_derivation}; therefore, the total mean EPR appears in 
\cref{eq:transition_probability_measure}. The path integral formulation of MinAP allows for the 
study of the stochastic optimal control problem \cite{Kappen_2005_prl, Kappen_2005_jstat}; 
however, by relying on the convergence of \cref{eq:transition_probability_measure} towards the 
most-likelyhood path of the path integral, we ignore the variance (stochasticity) associated with 
the EPR \cite{Kappen_2005_prl, Kappen_2005_jstat}, thereby focusing on the optimization of the 
mean EPR.
}
We exploit the MinAP framework to formulate and solve the generalized finite--time optimal 
control problem (GFTOC), see \cref{table:enrgertic_decomposition_EPR}.

\begin{table}[t!]
\centering
\begin{tabular}{c c c}
     \hline \hline
     \\ [-2.0ex]
     \textbf{Contribution to} & \textbf{Potential energetic} & \hspace{0.2cm} \textbf{Kinetic energetic} \\[-0.1ex]
     \vspace{1pt} \textbf{the action} & \textbf{dissipation} & \textbf{dissipation} \hspace{0.2cm} 
     \\ [0.2ex] 
     \hline
     \\ [-2.2ex]
     Thermodynamic & Nonquadratic TKUR & GFTOC \\ [-0.0ex]
     \vspace{1pt} law & and FR in Ref.\cite{atm_2025_var_epr_derivation} & in this work \\ [-0.2ex]
     \hline
     \\ [-2.2ex]
     {Variational} & & \\ [-0.0ex]
     \vspace{1pt} {optimization of} & \Cref{eq:lagrangian_full_contol_description} & \Cref{eq:optimal_driving_epr} \\ [0.1ex]
     \hline
     \\ [-2.2ex]
     {Control parameters} & & \\ [-0.0ex]
     \vspace{1pt} {of model are} & Fixed & Time--dependent \\ [0.1ex]
     \hline \hline
\end{tabular}
\caption{The nonquadratic TKUR and FR focus on the variational optimization of the \textit{potential-energetic} dissipation contribution to the action, with fixed control parameters, whereas GFTOC focuses on the variational optimization of the \textit{kinetic-energetic} dissipation contribution, with time-dependently varied control parameters. Unlike the \textit{potential-energetic} dissipation, the \textit{kinetic-energetic} dissipation depends on the path traversed in the control-parameter manifold.}
\label{table:enrgertic_decomposition_EPR}
\end{table}
\subsection{Summary}
\subsubsection{Optimization problem definition}
%

\begin{figure}[t!]
\centering
\includegraphics[width=\linewidth]{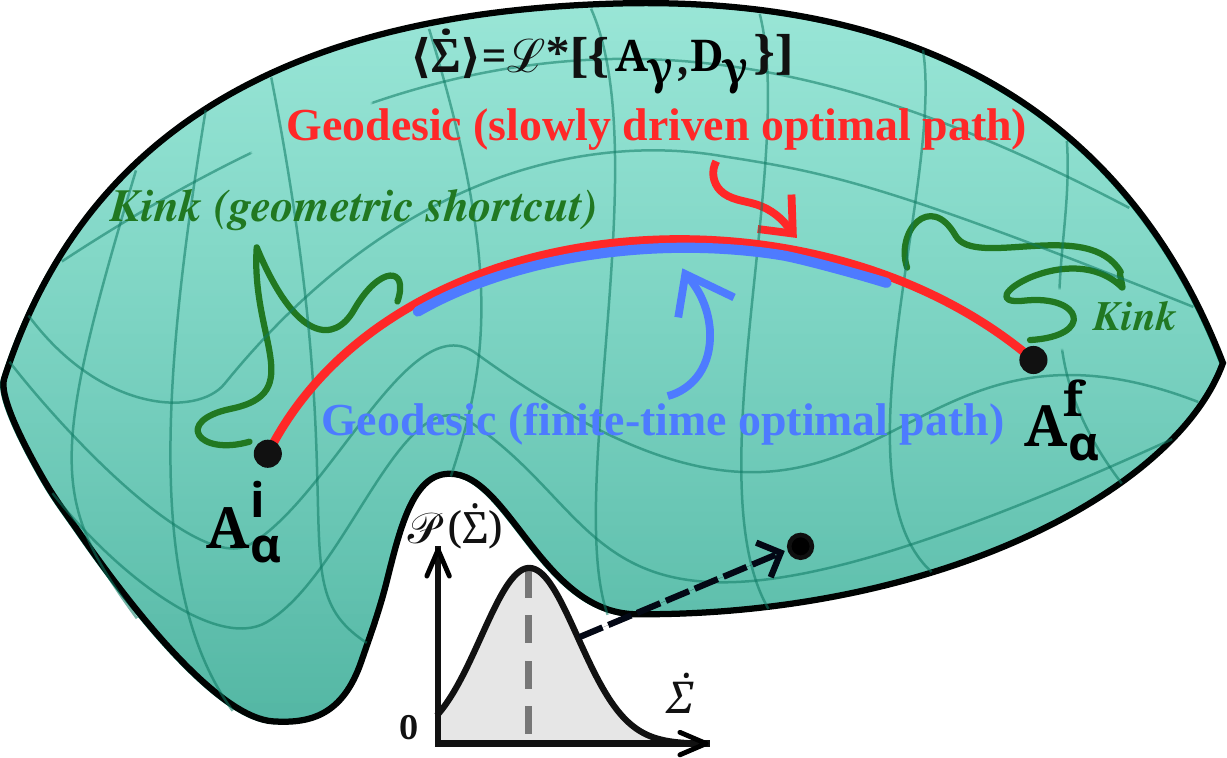}
\caption{ The Riemannian manifold of dissipation [$\langle\dot{\Sigma}\rangle$ in 
\cref{eq:lagrangian_full_contol_description}] is parameterized by the set $\{A_\gamma, D_\gamma\}$. Each point in the manifold has its respective probability distribution for stochastic EPR [\cref{eq:transition_probability_measure}] with mean $\langle\dot{\Sigma}\rangle$ [\cref{eq:lagrangian_full_contol_description}].
Here, $A_\alpha$ is the transition affinity driven optimally from an initial value $A_\alpha^i$ 
to a final value $A_\alpha^f$, a state-to-state transformation. As is evident, the \textit{curvature of the manifold} plays a key role in the driving process, the \textit{kinetic energetic} dissipation contribution is quantified by the driving EPR [\cref{eq:optimal_driving_epr}]. For a 
slowly driven system, the minimum-dissipation optimal path follows the geodesic on the manifold 
of thermodynamic dissipation; see the 
red line connecting $A_\alpha^i$ and $A_\alpha^f$. However, a finite driving time imposes a 
dynamical constraint, allowing only a sub-arc of the geodesic to be traversed within the given 
driving time; see the blue line. \textit{`Kinks'} are geometric thermodynamic shortcuts necessary for 
`swift state-to-state transformations'. }
\label{fig:illustration}
\end{figure}

{
We consider a generalized class of finite--time thermodynamic optimization problems (without the 
slow--driving assumption) termed GFTOC, defined as 
the minimization the \textit{kinetic energetic} term [the driving EPR \cref{eq:optimal_driving_epr}], required to drive the system from an initial control parameter $A_\alpha^i$ to a final control 
parameter $A_\alpha^f$ in a finite time $\tau$ \footnote{The driving time $\tau$ is different from the previously used observation time $\tau^{obs}$ used to delineate the short-time and long-time scaling symmetries of $\langle \dot{\Sigma} \rangle$.}, implemented through external parametric control 
of the affinity $A_\alpha \in \{A_{\alpha}\}$ on the manifold of $\langle\dot{\Sigma}\rangle$ [\cref{eq:lagrangian_full_contol_description}] an unavoidable \textit{potential energetic} term; see 
\cref{fig:illustration}. We fix the mobilities $D_\alpha$ (conjugate to $A_\alpha$) and focus 
on the optimal control of the affinities $A_\alpha$.
}

{
As evident from \cref{eq:mean_EPR}, three different classes of affinity control --- $A_\alpha \in 
\{\lambda_E\}$, $A_\alpha \in \{-\ln(\rho_i/\rho_i^E)\}$, and $A_\alpha \in \{F_\gamma\}$ --- 
correspond to the control and minimization of the free energy functional [$-\psi_E \propto O(1)$], excess 
EP [$\langle\Sigma_E^{ex}\rangle \propto O(1)$], and housekeeping EPR 
[$\langle\dot{\Sigma}^{hk}\rangle \propto O(1)$], respectively \footnote{As highlighted in 
\cref{sec:setup}, $\dot{\psi}_E$ and $\langle\dot{\Sigma}^{ex}\rangle$ are boundary terms in 
time; therefore, their optimization leads to a trivial solution. To address this, the optimization 
problem must be defined for $\psi_E$ and $\langle\Sigma^{ex}\rangle$. In contrast, 
$\langle\dot{\Sigma}^{hk}\rangle$ cannot be integrated out in time; therefore, the optimization 
problem must be defined for $\langle\dot{\Sigma}^{hk}\rangle$.}.
Importantly, since the control parameters are driven in a finite time $\tau$, we define the 
optimization problem for a thermodynamic dissipation term that admits $O(1)$ scaling. 
\textit{Class~(1):} $A_\alpha \in 
\{\lambda_E\}$, relaxed systems that satisfy the state--space Boltzmann distribution (in 
equilibrium or at a non-equilibrium steady state) at all times are fully characterized by 
$\psi_E$. Loosely speaking, the slow--driving formulation of thermodynamic geometry belongs 
to this class. 
\textit{Class~(2):} $A_\alpha \in 
\{-\ln(\rho_i/\rho_i^E)\}$, systems that lack relaxation to the state--space Boltzmann distribution 
require control of the excess EP $\langle\Sigma^{ex}\rangle$, achieved through explicit 
state--space control. Loosely speaking, the slow--driving formulation of optimal transport theory belongs to this class. 
\textit{Class~(3):} $A_\alpha 
\in \{F_\gamma\}$, as highlighted in \cref{sec:motivation,sec:setup}, the optimal control of 
the steady-state dissipation of the graph $\langle\dot{\Sigma}^{hk}\rangle$ is entirely absent; 
it lies in the transition space \cite{atm_2024_var_epr, atm_2025_var_epr_derivation} and 
therefore dominates in the long-time limit, when the NESS satisfies the state--space Boltzmann 
distribution due to relaxation.
}
%

{
Since the set $\{A_\alpha\} = \{\{\lambda_E\}, \{-\ln(\rho_i/\rho_i^E)\}, \{F_\gamma\}\}$ of all 
possible transition affinities is linearly independent, it suffices to demonstrate the underlying 
physical mechanism of GFTOC using a scalar control parameter affinity $A_\alpha$; the 
generalization to multi--parameter GFTOC follows trivially in \cref{sec:optimal_control_generalization}, 
extending to both linearly independent and dependent controllable affinities, as well as the 
simultaneous optimal control of affinities and mobilities $\{A_\alpha, D_\alpha\}$, with conjugate mobilities $\{ D_\alpha \} = \{\{1\}, \{1\},\{T_\gamma^\perp\}\}$, as it is evident from \cref{eq:mean_EPR}.
}
\subsubsection{Overview of main results}
\begin{table}[t!]
\centering
\begin{tabular}{c c c c}
     \hline \hline
     \\ [-2.0ex]
     \textbf{GFTOC} \: & {Potential} & \hspace{0.2cm} {Kinetic} & Application 
     \\ [0.2ex] 
     & {energetic} & \hspace{0.2cm} {energetic} & detailed in 
     \\ [0.2ex] 
     \hline
     \\ [-2.0ex]
     \textbf{\textit{Class~(1)}} \: & $-\Delta_{\lambda_E} \psi_E$ & $\frac 1 2 \sum_{\{\lambda_E\}} \int_0^{\tau} \dot{\lambda_E}^2 \partial_{\lambda_E}^2 \psi_E \: dt$ & \Cref{sec:free_energy} \\ [1.2ex]
     \hline
      &  &  & \Cref{sec:harmonic_trap} \\ [0.0ex]
     \hline
     \\ [-2.0ex]
     \textbf{\textit{Class~(2)}} \: & $-\Delta_{\lambda_E} D_E^{KL}$ 
     & $\frac 1 2 \sum_{\{i\}} \int_0^{\tau} [{\partial_t \ln{(\rho_i/\rho_i^E)}}]^2 \rho_i \: dt$
     & \Cref{sec:excess_epr} \\ [1.2ex]
     \hline
      &  &  & \Cref{sec:linear_optimal_control} \\ [0.0ex]
     \hline
     \\ [-2.0ex]
     \textbf{\textit{Class~(3)}} \:& $ \langle \dot{\Sigma}^{hk} \rangle$ & $\frac 1 2 \sum_{\{\gamma^\rightleftharpoons\}} \int_0^{\tau} \dot{F}_\gamma^2 \: \partial_{F_\gamma}^2 \langle \dot{\Sigma}^{hk} \rangle \: dt$
     & \Cref{sec:housekeeping_epr} \\ [1.2ex]
     \hline \hline
\end{tabular}
\caption{ Thermodynamic geometry and optimal transport theory belong to the slow-driving formulation of 
\textit{Class~(1)} and \textit{Class~(2)}, respectively; see \cref{fig:overview}. The 
connection of \textit{Class~(2)} to optimal transport theory becomes evident after applying the 
conditions $D_E^{KL}(\vec{\lambda}^{inl}) = 0$ (initially equilibrated state distribution) and 
$D_E^{KL}(\vec{\lambda}^{fnl}) \neq 0$ (penalty cost of the target state distribution) for the 
\textit{potential-energetic} contribution, and substituting $\partial_t\ln(\rho_i/\rho_i^E) = 
v_i$ (allowed by the continuity equation) together with the discrete-to-continuous state--space 
change $(\sum_{\{i\}} \to \int_i)$ for the \textit{kinetic-energetic} contribution; the Fisher 
information then reduces to the transportation cost $\int_i\int_0^\tau\rho_i v_i^2\,dt$. Here, 
$v_i$ is the solution of the continuity equation; therefore, $v_i$ incorporates non-vanishing non-equilibrium 
currents. Hence, optimal transport theory is a cEQ `hybrid' case that accounts for a 
quadratic dissipation cost but uses the `\textit{curvature}' of excess EP. \Cref{sec:harmonic_trap} connects thermodynamic geometry and 
optimal transport theory. Similarly, \cref{sec:linear_optimal_control} elucidates the 
connection between \textit{Class~(2)} and \textit{Class~(3)}.
}
\label{table:GFTOC_three_classes}
\end{table}
Here, based on MinAP \cite{atm_2024_var_epr, atm_2025_var_epr_derivation}, we formulate the 
GFTOC framework, valid for any discrete-state fEQ system, by exploiting the Riemannian geometric 
structure in the control parameter space. We compute the optimal finite--time driving protocols 
that minimize the total driving EP, which exhibit \textit{kinks}. GFTOC unifies the optimal 
control of slowly driven and finite--time processes, as we prove an exact geometric mapping 
between them. Importantly, since MinAP assigns a thermodynamic cost to any transition-path 
realization, a thermodynamic cost for sustaining the \textit{kinks} is quantified (a boundary 
term in driving time), namely the thermodynamic shock. Thereby, the \textit{kinks} reduce the 
thermodynamic cost of driving (a bulk term in driving time), a `geometric shortcut', since the 
\textit{kinks} reduce the distance to be travelled in the control parameter space. The 
thermodynamic interplay between the boundary and bulk terms of driving EP gives rise to the 
formation of \textit{kinks} as a dynamical consequence, and the restoration of `the finite--time 
geometric speed limit', an inherent physical timescale associated with the optimal driving 
process --- `a time--dissipation--geometry tradeoff relation'. The dual manifestation of the 
bulk--boundary terms of driving EP has a physical interpretation analogous to work--heat, but 
formulated here for the finite--time optimal driven process in ST;
{see \cref{fig:illustration}}.

\begin{figure*}[t!]
\centering
\includegraphics[width=\textwidth]{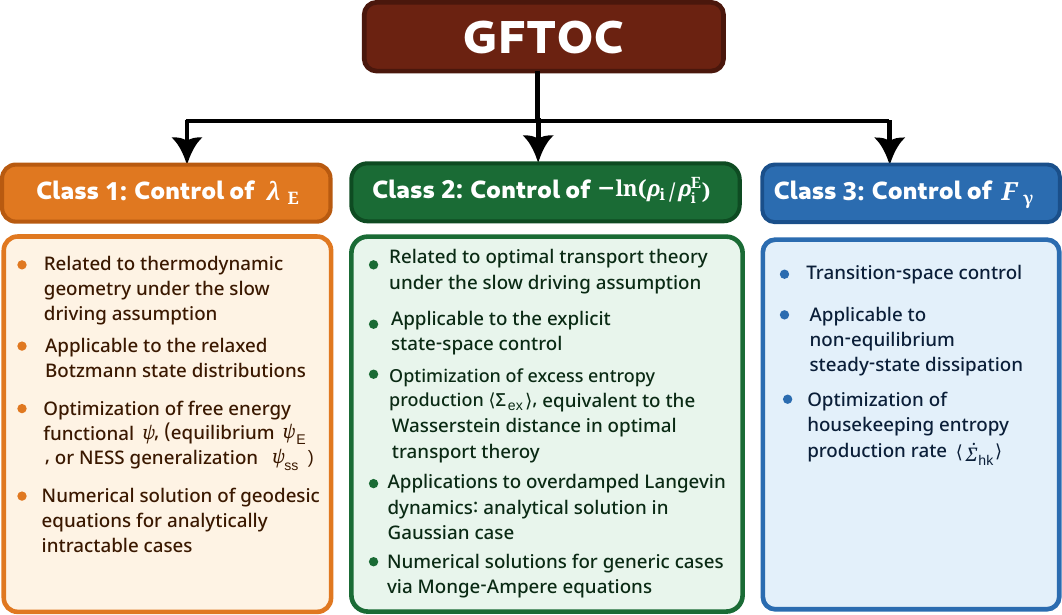}
\caption{GFTOC systematically incorporates the existing close-to-equilibrium and slow--driving 
methodologies of thermodynamic geometry and optimal transport theory. GFTOC formulates a 
far-from-equilibrium and finite--time driving of any control parameters on a Riemannian geometric 
manifold of total dissipation $\langle\dot{\Sigma}\rangle$; see \cref{fig:illustration}. The 
threefold orthogonal decomposition of dissipation [\cref{eq:mean_EPR}] reduces GFTOC to three equivalence 
classes.}
\label{fig:overview}
\end{figure*}

We demonstrate the applicability of GFTOC through several examples and address consistency 
issues arising from existing model-specific optimal control formulations. This leads to a unified 
and universal theoretical framework for optimal control in ST. Our approach provides a threefold 
realization of the minimum action principle (MinAP) in ST via the non-quadratic TKUR, fluctuation 
relations (FR), and GFTOC, corresponding to thermodynamic inference, partial control, and full 
control descriptions, respectively \cite{atm_2024_var_epr}. The full control description enabled 
by GFTOC allows for the thermodynamically efficient design and control of finite-size stochastic 
systems in finite driving time, where ST has been experimentally validated. The 
stochastic-thermodynamic insights of \textit{kinks} in GFTOC advance optimal transport theory 
and thermodynamic geometry beyond the slow--driving and close-to-equilibrium assumptions;
{illustrated in \cref{fig:overview} and the corresponding optimal control problems are summarized in \cref{table:GFTOC_three_classes}.}
Since information geometry accounts only for excess EP and lacks \textit{kinks} 
\cite{Rao_1945,Amari_2016_book}, GFTOC further advances the state of the art in information geometry.

\subsubsection{Technical outline of main results}
We explicitly identify five \textit{shortcomings} of the existing cEQ optimal 
control methodologies implemented in ST so far.
\begin{itemize}
    \item[{(S1):}] Violation of the Riemannian geometric structure in the control parameter space.
    \item[{(S2):}] Exclusion of the housekeeping EPR.
    \item[{(S3):}] Use of quadratic cEQ dissipation functions for the EPR.
    \item[{(S4):}] Constraint of finite driving time.
    \item[{(S5):}] Lack of a thermodynamic and dynamical origin and interpretation of \textit{kinks}.
\end{itemize}
The exact model-specific results in Ref.~\cite{Schmiedl_2007} focus on resolving $(\text{S}4)$ 
but do not address $(\text{S}1)$, $(\text{S}2)$, or $(\text{S}5)$. Despite `formally' addressing 
$(\text{S}2)$, optimal transport theory is prone to $(\text{S}1)$, $(\text{S}3)$, and 
$(\text{S}4)$\footnote{Optimal transport theory introduces an incremental development regarding 
$(\text{S}3)$ that addresses $(\text{S}2)$; however, it remains a fundamental shortcoming for 
fEQ systems, see Refs.~\cite{atm_2024_var_epr, atm_2025_var_epr_derivation}. Therefore, by 
construction, methodologies with $(\text{S}2)$ are also prone to $(\text{S}3)$.}. Thermodynamic 
geometry aims to address $(\text{S}1)$ but remains susceptible to $(\text{S}2)$ and 
$(\text{S}4)$\footnote{By construction, methodologies with $(\text{S}4)$ are also prone to 
$(\text{S}5)$.}.

The key results of our framework are:%
\begin{itemize}
    \item[{\Hygiea}]
    \Cref{sec:optimal_slow_driving} addresses shortcomings $(\text{S}1)$, $(\text{S}2)$, and 
    $(\text{S}3)$ under the slow--driving assumption. The class of optimal fEQ driving protocols 
    given by universal geodesics [\cref{eq:geodesic_expressions}] is defined as the linear 
    interpolation in the geodesic space of the control parameters [\cref{eq:optimal_protocol}]. 
    The corresponding dynamical, thermodynamic, and timescale quantities [\cref{eq:optimal_protocol}, 
    \cref{eq:optimal_driving_EP}, and \cref{eq:slow_driving_speed_limit}] are defined as the 
    optimal driving protocol, optimal driving EP, and geometric speed limit, respectively, 
    quantifying the minimal driving distance, dissipation, and time.

    \item[{\Hygiea}]
    \Cref{sec:finite_time_optimal_control} addresses shortcomings $(\text{S}4)$ and $(\text{S}5)$ 
    via an exact mapping to the optimal slow--driving control problem in 
    \cref{sec:optimal_slow_driving}. The resulting finite--time dynamical, thermodynamic, and 
    timescale counterparts are given by \cref{eq:optimal_geodesic_finite_time_explicit}, 
    \cref{eq:optimal_finite_time_total_EP}, and \cref{eq:slow_driving_speed_limit_finite_time}. 
    The thermodynamic cost [\cref{eq:optimal_finite_time_bulk_boundary_EP}] admits a 
    bulk--boundary decomposition, analogous to work and heat, which is physically associated 
    with slow driving and the generation of \textit{kinks}, respectively. Thereby, the 
    finite--time geometric speed limit is restored [\cref{eq:slow_driving_speed_limit_finite_time}]: 
    `a time--dissipation--geometry tradeoff relation'.

    \item[{\Hygiea}]
    \Cref{sec:optimal_control_generalization} generalizes the analytical framework to two 
    practically applicable settings: multiple control parameters and novel numerical algorithms. 
    \Cref{sec:application_GFTOC} applies the framework to case-specific problems, highlighting 
    novel results and solving several previously open problems.
\end{itemize}
\section{Slow Driving Optimal Control}\label{sec:optimal_slow_driving}
\subsection{Geodesic structure}\label{sec:geodesic_structure}
We consider slow driving of the Lagrangian ($\mathcal{L}^*$), which physically implies that 
the system dynamics and statistics quickly adapt to the instantaneous control parameters. Hence, 
the slow--driving formulation is valid under a timescale separation: precisely, only if the 
driving timescale is slower than the coupling timescale between the environment and the system, 
or the relaxation timescale of the system, which quantifies how quickly the system adapts to 
the instantaneously imposed control parameters. Under the assumption of slow driving, we expand 
the Lagrangian to second-order terms in the rate of driving $\dot{A}_\alpha$. The driving 
Lagrangian reads \cite{Salamon_1985, Brody_1995, Schlogl_1985, Crooks_2007}:
\begin{equation}\label{eq:optimal_driving_epr}
\begin{split}
    \mathcal{L}_{drv}^*\left[A_{\alpha}, \dot{A}_\alpha\right] = \frac{1}{2} \partial_{{\alpha}}^2 \mathcal{L}^* \left(\dot{A}_{{\alpha}}\right)^2,
\end{split}    
\end{equation}
such that the total driving EP is $\mathcal{S}_{drv}^{qs} = \Sigma_{drv}^{qs} = \int_0^\tau \mathcal{L}_{drv}^* \, dt$.

\begin{figure}[t!]
\centering
\includegraphics[width=\linewidth]{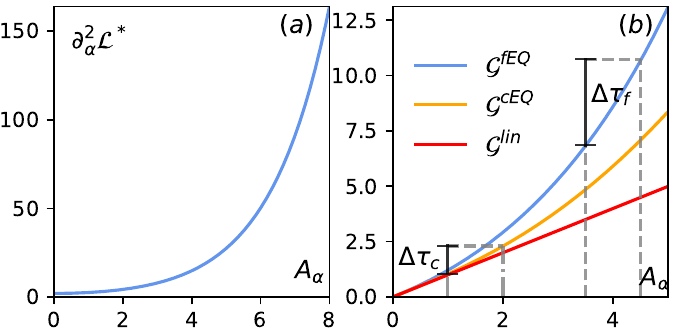}\label{fig:sub1}
\caption{(a) The curvature of the Lagrangian $\partial_\alpha^2 \mathcal{L}^*$ (\textit{mass}  given by 
\cref{eq:driving_mass}) increases exponentially with the driving affinity $A_\alpha$. (b) 
$\mathcal{G}(A_\alpha): A_\alpha \to t$: comparison between $\mathcal{G}^{cEQ}(A_\alpha)$, 
$\mathcal{G}^{fEQ}(A_\alpha)$, and $\mathcal{G}^{lin}(A_\alpha)$ given by 
\cref{eq:geodesic_expressions}. For fixed quasi-static driving speed $v_{qs}$, 
$A_\alpha^f - A_\alpha^i = 1$ is considered for cEQ ($A_\alpha^i = 1$) and fEQ 
($A_\alpha^i = 3.5$), and the corresponding $\Delta\tau_c$ and $\Delta\tau_f$ are plotted; 
this is a combined pictorial visualization of 
\cref{eq:geodesic_expressions,eq:slow_driving_speed_limit}. Due to 
\cref{eq:slow_driving_speed_limit}, the geometric speed limit for quasi-static slow driving 
increases for fEQ systems, an effect attributed to the higher \textit{mass} required for driving fEQ 
systems.}
\label{fig:1}
\end{figure}

\Cref{eq:optimal_driving_epr} quantifies the `\textit{driving kinetic energy}', where the \textit{mass} is the instantaneous `\textit{local curvature}' of the Lagrangian 
($\partial_{{\alpha}}^2 \mathcal{L}^*$) \cite{Salamon_1985, Brody_1995, Schlogl_1985, 
Crooks_2007}. We adopt the shorthand notation $\partial_{{\alpha}}^2 \mathcal{L}^* = 
\partial_{A_{\alpha}}^2 \mathcal{L}^*$, which satisfies
\begin{equation}\label{eq:driving_mass}
    \partial_{{\alpha}}^2 \mathcal{L}^* = D_{\alpha} \left[ 2\cosh\!\left(\frac{A_{\alpha}}{2}\right) + \frac{1}{2} A_\alpha \sinh\!\left(\frac{A_{\alpha}}{2}\right) \right] = T_\alpha + \frac{1}{4} \mathcal{L}^*
\end{equation}
Far-from-equilibrium driven systems (with larger $A_\alpha$) exhibit a higher mass, attributed 
to larger fluctuations ($T_\alpha$) and dissipation ($\mathcal{L}^*$). Physically, this 
signifies a higher resistance to driving in fEQ systems; \cref{fig:1}\textcolor{red}{(a)} plots 
\cref{eq:driving_mass}. The fEQ driving is critically slowed due to increased resistance, with 
a singularity in the $A_\alpha \to \infty$ limit caused by the divergence of the EPR and 
traffic. Importantly, the local curvature $\partial_{{\alpha}}^2 \mathcal{L}^*$ generalizes 
the stochastic Fisher information for any $\mathcal{L}^*$, thereby going beyond the existing 
formulations in information geometry \cite{Rao_1945,Amari_2016_book} and thermodynamic geometry 
\cite{Salamon_1985, Brody_1995, Schlogl_1985, Crooks_2007}. Choosing the excess EP as the 
total dissipation cost reveals the equivalence between the \textit{mass}  and the stochastic Fisher 
information for each state, which is the predominantly studied dissipation function in 
information geometry \cite{Rao_1945,Amari_2016_book}, discussed further in \cref{sec:excess_epr}.
{
Importantly, \cref{eq:driving_mass} incorporates the impact of both non-equilibrium traffic 
(scaled fluctuations) and dissipation, in contrast to thermodynamic geometry 
\cite{Salamon_1985, Brody_1995, Schlogl_1985, Crooks_2007} and information geometry 
\cite{Rao_1945,Amari_2016_book}, which only account for equilibrium and non-equilibrium traffic (scaled fluctuations), 
respectively.
}

We have omitted the \textit{`potential energy'} boundary term, namely the instantaneous mean EPR [\cref{eq:lagrangian_full_contol_description}], which does 
not depend on driving speed, time or the path of control parameter change, and corresponds to the unavoidable instantaneous minimum 
EPR cost in the limit $\tau \to \infty$. Since the \textit{`driving kinetic energy'} depends on the path of control parameter change, the slow--driving optimization problem 
reduces to minimizing the \textit{`driving kinetic energy'} (driving EPR [\cref{eq:optimal_driving_epr}]). Its solution is the 
geodesic equation for $A_\alpha$,
\begin{equation}\label{eq:geodesic_equation}
\begin{split}
    \ddot{A}_\alpha + \frac{\partial_{{\alpha}}^3 \mathcal{L}^*}{2\partial_{{\alpha}}^2 \mathcal{L}^*} \dot{A}_\alpha^2 = 0,
\end{split}    
\end{equation}
\Cref{eq:geodesic_equation} generalizes the geodesic equation for the optimal control of fEQ 
systems that minimizes the driving EPR \cite{Salamon_1985, Brody_1995, Schlogl_1985, Crooks_2007, Sekimoto_1997_slow_driving_optimal_control, 
Sivak_2012}. It equivalently implies $\dot{A}_\alpha \sqrt{\partial_{{\alpha}}^2 \mathcal{L}^*} 
= v_{qs}$, where $v_{qs}$ is the quasi-static driving speed along the geodesic in driving time 
$\tau$. The minimum action solution for optimal driving is called the geodesic and is denoted 
by $\mathcal{G}(A_\alpha)$. It encodes the minimum-distance path in the control parameter space, 
and its analytical form plays a crucial role. By definition, the geodesic satisfies 
$\mathcal{G}(A_\alpha) = \arg\inf_{A_\alpha}[\Sigma_{drv}^{qs}]$, and the 
corresponding minimum EP is $\Sigma_{qs}^* = \inf_{A_\alpha}[\Sigma_{drv}^{qs}]$, 
under the initial and final boundary conditions $A_\alpha^i$ and $A_\alpha^f$. Thus, 
$\dot{A}_\alpha \propto 1/\sqrt{\partial_{{\alpha}}^2 \mathcal{L}^*}$ implies that $A_\alpha$ 
must be driven more slowly the further the system is from equilibrium. Physically, this means 
that for the optimal driving to be valid, each infinitesimally small driving timestep $\Delta t$ 
contributes equally to $\Sigma_{drv}^{qs}$.

Computing the geodesic gives the mapping between driving speed, time, and control parameter, 
$A_{\alpha}(t) = \mathcal{G}^{-1}(v_{qs} t)$. Here, $\mathcal{G}(A_\alpha): A_\alpha \to t$ 
is a function mapping the instantaneous control parameter to time, and its inverse is 
$\mathcal{G}^{-1}(v_{qs} t): t \to A_\alpha$. Thus, in geodesic space, the driving protocol 
satisfies the linear solution $\mathcal{G}(A_\alpha) = c_0 + v_{qs} t$, with $c_0$ and $v_{qs}$ 
being unknowns to be determined. To this end, slow driving inherently imposes boundary conditions 
at the initial and final times. Hence, $\mathcal{G}(A_\alpha^i) = c_0$ and 
$\mathcal{G}(A_\alpha^f) = v_{qs} \tau + c_0$, which imply $c_0 = \mathcal{G}(A_\alpha^i)$ and 
$v_{qs} = [\mathcal{G}(A_\alpha^f) - \mathcal{G}(A_\alpha^i)]/\tau$. This reduces the optimal 
driving protocol to a linear interpolation in geodesic space between the initial and final 
control parameters:
\begin{equation}\label{eq:optimal_protocol}
\begin{split}
    \mathcal{G}(A_\alpha) = \left(1 - \frac{t}{\tau}\right)\mathcal{G}(A_\alpha^i) + \frac{t}{\tau}\mathcal{G}(A_\alpha^f).
\end{split}    
\end{equation}
Hence, $v_{qs} = \partial_t \mathcal{G}(A_\alpha) = \frac{1}{\tau}\left[\mathcal{G}(A_\alpha^f) 
- \mathcal{G}(A_\alpha^i)\right]$. This simplifies to $\Sigma_{qs}^* = \inf_{A_\alpha} 
\left(\Sigma_{drv}^{qs}\right) = \inf_{A_\alpha}\int_0^{\tau}\mathcal{L}_{drv}^* \, dt = 
\frac{1}{2}\tau(v_{qs})^2 = \frac{1}{2}\tau\left(\partial_t \mathcal{G}(A_\alpha)\right)^2$. 
Thus, the minimum slow--driving EP cost reads
\begin{equation}\label{eq:optimal_driving_EP}
\begin{split}
    \Sigma_{qs}^* = \frac{1}{2\tau}\left[\mathcal{G}(A_\alpha^f) - \mathcal{G}(A_{\alpha}^i)\right]^2.
\end{split}    
\end{equation}
This simplification allows us to visualize the physical meaning of geodesic space using 
\cref{eq:optimal_protocol,eq:optimal_driving_EP}. In particular, the optimal control problem 
for $A_\alpha$ with a varying instantaneous local \textit{mass} is converted into one with a unit \textit{mass} using the geodesic map $\mathcal{G}(A_\alpha)$. Similarly, time is scaled in units of the 
driving time $\tau$. This mapping converts the non-quadratic optimal control problem in 
$A_\alpha$ space into a quadratic optimal control problem in $\mathcal{G}(A_\alpha)$ space, 
with a scaled unit \textit{mass} and a scaled unit time. This is a fundamental non-equilibrium 
scale-invariance relation for the optimal quasi-static driving of fEQ systems, and it 
delineates the underlying fundamental universality of all quasi-statically driven systems.
\subsection{\label{sec:geodesic_expressions}Exact analytical expressions for geodesics}
In this section, we compute exact analytical expressions for the geodesic for the control of 
the transition affinity $A_\alpha$ for three different cases. We fix the amplitude of the \textit{mass} to $1$ in the $A_\alpha \to 0$ limit. The first case is the linear geodesic $\mathcal{G}^{lin}$, 
computed using the equilibrium approximation of the \textit{mass} $\partial_{{A_\alpha}}^2 \mathcal{L}^* 
= 1$, corresponding to Euclidean geometry. The second case is the close-to-equilibrium geodesic $\mathcal{G}^{cEQ}$, computed using 
the close-to-equilibrium approximation: up to $O(A_{\alpha}^2)$ terms of 
$\partial_{{A_\alpha}}^2 \mathcal{L}^*$, this implies $\partial_{{A_\alpha}}^2 \mathcal{L}^* = 
1 + A_\alpha^2/4$. The third case is the fEQ geodesic $\mathcal{G}^{fEQ}$, computed using the 
fEQ approximation $\cosh(a) \approx \sinh(a) \approx \frac{1}{2}e^a$, implying 
$\partial_{{A_\alpha}}^2 \mathcal{L}^* = \frac{1}{4}(A_\alpha + 4)e^{\frac{1}{2}A_\alpha}$.

The geodesic is obtained by solving the ODE $\dot{A}_\alpha\sqrt{\partial_{{A_\alpha}}^2 
\mathcal{L}^*} = v_{qs}$. In addition, we impose the constraint $\mathcal{G}(0) = 0$, 
corresponding to fixing the integration constant, which allows comparison between different 
geodesics from the same reference point. The exact closed-form analytical expressions for the 
linear geodesic $\mathcal{G}^{lin}$, close-to-equilibrium geodesic $\mathcal{G}^{cEQ}(A_\alpha)$, 
and far-from-equilibrium geodesic $\mathcal{G}^{fEQ}(A_\alpha)$ are:
\begin{equation}\label{eq:geodesic_expressions}
\begin{split}
    \mathcal{G}^{lin}(A_\alpha) & = A_\alpha,
    \\
    \mathcal{G}^{cEQ}(A_\alpha) & = \frac{1}{4}A_\alpha\sqrt{4+A_\alpha^2} + \sinh^{-1}\!\left(\frac{A_\alpha}{2}\right),
    \\
    \mathcal{G}^{fEQ}(A_\alpha) & = 2\sqrt{\left(4 + A_\alpha\right)e^{\frac{1}{2}A_\alpha}} - \frac{2\sqrt{\pi}}{e}\,\mathrm{ErfI}\!\left[\frac{1}{2}\sqrt{4 + A_\alpha}\right]
    \\ & \hspace{0.5cm} - 4 + \frac{2\sqrt{\pi}}{e}\,\mathrm{ErfI}[1],
\end{split}    
\end{equation}
respectively, where $\mathrm{ErfI}[x]$ is the imaginary error function. The analytical form of 
$\mathcal{G}(A_\alpha)$ plays a key role. $\mathcal{G}^{fEQ}$, $\mathcal{G}^{cEQ}$, and 
$\mathcal{G}^{lin}$ are plotted in \cref{fig:1}\textcolor{red}{(b)}. The geodesics satisfy the 
hierarchy of inequalities,
\begin{equation}\label{eq:geodesic_hierarchy}
\begin{split}
    \mathcal{G}^{fEQ}(A_\alpha) \geq \mathcal{G}^{cEQ}(A_\alpha) \geq \mathcal{G}^{lin}(A_\alpha).
\end{split}    
\end{equation}
In the $A_\alpha \to 0$ limit, $\mathcal{G}^{lin}(A_\alpha)$, $\mathcal{G}^{cEQ}(A_\alpha)$, 
and $\mathcal{G}^{fEQ}(A_\alpha)$ converge. However, the further the system is from 
equilibrium, the greater the quantitative difference observed. The physical implication is a 
tighter bound on the exact optimal driving EP in \cref{eq:optimal_driving_EP}, and a deviation 
of the optimal driving protocol \cref{eq:optimal_protocol} from a trivial linear interpolation 
between the initial and final control parameters. This highlights the importance of the exact 
fEQ geodesics.
\subsection{Far--from--equilibrium generalizations of \newline optimal transport theory}
The geodesic hierarchy \cref{eq:geodesic_hierarchy} implies the hierarchy 
$[\mathcal{G}^{fEQ}(A_\alpha^f) - \mathcal{G}^{fEQ}(A_{\alpha}^i)]^2 \geq 
[\mathcal{G}^{cEQ}(A_\alpha^f) - \mathcal{G}^{cEQ}(A_{\alpha}^i)]^2 \geq 
[A_\alpha^f - A_{\alpha}^i]^2$; see \cref{fig:1}\textcolor{red}{(b)}. For a fixed $\tau$, 
the hierarchy on the driving EP obtained using \cref{eq:optimal_driving_EP} reads,
\begin{equation}\label{eq:driving_EP_hierarchy}
\begin{split}
    \Sigma_{qs}^{*fEQ} \geq \Sigma_{qs}^{*cEQ} \geq \Sigma_{qs}^{*lin}.
\end{split}    
\end{equation}
$\mathcal{G}^{lin}(A_\alpha)$ is the special case that has been employed in optimal transport 
theory \cite{Jordan_1998, Benamou_2000} and applied to ST to obtain $\Sigma_{qs}^{*lin}$ as 
the driving EP cost, which is a quadratic function of the change in the driving affinity 
\cite{Aurell_2011, Van_vu_2023}. In this respect, optimal transport theory is a cEQ `hybrid' 
case that operates in a `state--space' representation, for which the \textit{mass} for the excess EP is 
employed (see \cref{sec:excess_epr}) \footnote{There is a minor technical difference that is not 
of fundamental importance, since the \textit{mass} defined for a control parameter is the inverse of the 
\textit{mass} defined for its dual conjugate; see \cref{sec:excess_epr}.}, but a cEQ quadratic dependence 
on the driving affinity incorporates the housekeeping EPR (see \cref{sec:linear_optimal_control}).

However, the computation of $\mathcal{G}$ takes into account the non-quadratic nature of the 
dissipation function $\mathcal{L}^*$, which yields the tightest and exact bound using 
\cref{eq:driving_EP_hierarchy}. Moreover, the right-hand side of \cref{eq:optimal_driving_EP} 
is equal to the square of the $\mathcal{W}_2$ Wasserstein distance, defined as
\begin{equation}\label{eq:waseerstein_distacne}
\begin{split}
    \mathcal{W}_2 = \mathcal{G}(A_\alpha^f) - \mathcal{G}(A_{\alpha}^i),
\end{split}    
\end{equation}
the distance in the geodesic space of the control parameters. Recalling the definition of the 
geodesic, $\mathcal{W}_2$ physically corresponds to the minimization of distance and is 
mathematically equal to the $L^2$-norm. The definition \cref{eq:waseerstein_distacne} of 
$\mathcal{W}_2$ is itself novel and accounts for the geometric curvature of any control 
parameter. The existing cEQ optimal transport theory focuses on the control of the state space. 
Thus, \cref{eq:optimal_driving_EP} yields an equality between the driving EP in ST and the 
Wasserstein distance in optimal transport theory,
\begin{equation}
    \Sigma_{qs}^* = \frac{\mathcal{W}_2^2}{2\tau}.
\end{equation}
Therefore, our optimal control formulation goes beyond the existing quadratic cEQ counterparts 
formulated using optimal transport theory \cite{Aurell_2011, Van_vu_2023, Ito_2024_omtp_st, 
Oikawa_2025}. Importantly, our slow--driving formulation also reveals the connection between 
`thermodynamic geometry' and `optimal transport theory' beyond known quadratic formulations, 
extending the applicability of `optimal transport theory' to any non-Euclidean (Riemannian) 
control parameter manifolds.  
\subsection{Geometric speed limit for slow driving: \newline time--dissipation--geometry tradeoff relation}
The validity of slow driving is a key assumption in the optimal control framework developed so 
far. \Cref{eq:optimal_driving_EP} reveals an inherent timescale associated with the driving 
process. Inverting \cref{eq:optimal_driving_EP}, we define the quasi-static driving timescale:
\begin{equation}\label{eq:slow_driving_speed_limit}
\begin{split}
    \tau_{qs}^* = \frac{\left[\mathcal{G}(A_\alpha^f) - \mathcal{G}(A_{\alpha}^i)\right]^2}{2\Sigma_{qs}^*}.
\end{split}    
\end{equation}
$\tau_{qs}^*$ is a quantitative measure of the time required for given initial and final control 
parameters ($A_\alpha^i$ and $A_\alpha^f$) and driving EP $\Sigma_{qs}^*$. It quantifies the 
timescale for the violation of the slow--driving assumption. Physically, it is equal to the 
square of the distance travelled along the geodesic in the control parameter space, divided by 
the quasi-static EP supplied for the driving process.

The quasi-static driving time reveals the fundamental tradeoff between driving time, dissipation 
(EP), and geometry for the optimally driven process, and defines the geometric speed limit for 
the slow--driving process. Therefore, \cref{eq:slow_driving_speed_limit} is a fundamental 
`time--dissipation--geometry tradeoff relation'. Importantly, the slow--driving assumption is 
violated for any $\tau$ and does not necessarily require $\tau \leq \tau_{qs}^*$. For a fixed 
value $\Delta A_\alpha = A_\alpha^f - A_\alpha^i$ and a fixed available budget for driving EP 
$\Sigma_{qs}^*$, \cref{eq:geodesic_hierarchy} implies,
\begin{equation}\label{eq:timescale_hirerarchy}
    \tau_{qs}^{*fEQ} > \tau_{qs}^{*cEQ} > \tau_{qs}^{*lin},
\end{equation}
see \cref{fig:1}\textcolor{red}{(b)}. Hence, fEQ systems require a larger quasi-static driving 
time, attributed to critical slowing due to higher mass. This implies that fEQ systems are more 
prone to violating the slow--driving assumption, with prominent consequences discussed in 
\cref{sec:finite_time_optimal_control}.
\section{Finite--Time Optimal Control}\label{sec:finite_time_optimal_control}
The finite--time optimal protocol exhibits discontinuous jumps at the endpoints of the protocol, 
namely \textit{kinks} \cite{Schmiedl_2007, Aurell_2011, Chen_2019_stochastic_control, 
Zhong_2024}. In contrast, the slow--driving approach misses such jumps 
\cite{Sekimoto_1997_slow_driving_optimal_control, Sivak_2012}. The optimal slow--driving control 
framework developed in \cref{sec:optimal_slow_driving} relies on the key assumption of a 
timescale separation between the driving time $\tau$ and the largest inherent relaxation 
timescale of the system. This implies that the system relaxes instantaneously to the control 
parameter dictated by the environment, justifying the quasi-static slow--driving assumption. 
However, finite--time optimal processes may operate on shorter timescales such that 
$\tau < \tau_{qs}^*$, violating the slow--driving assumption. Moreover, the slow--driving 
assumption is also violated for $\tau \geq \tau_{qs}^*$, except in the $\tau \to \infty$ limit. 
To this end, in this section we develop the finite--time optimal control framework, combining 
MinAP with the slow--driving optimal control and the geodesic structure.

\subsection{Finite--time geodesic structure}\label{sec:finite_time_geodesic}
In a finite--time driving process, the system realizes that it cannot quasi-statically follow 
the geodesic. The GFTOC formulation therefore implies that the boundary condition constraint on 
the geodesic must be relaxed and treated as an optimization parameter. We consider 
$\mathcal{G}(A_{\alpha}^{i*})$ and $\mathcal{G}(A_{\alpha}^{f*})$ as the optimal control 
parameter values in the geodesic space at $t = 0^+ > 0$ and $t = \tau^- < \tau$, respectively, 
where $0^+$ and $\tau^-$ are infinitesimal times after the initial and before the final time, 
respectively. Hence, the protocol jumps in the geodesic space at the initial and final times are 
$\Delta_{0+}\mathcal{G}_{\tau}(A_\alpha) = \mathcal{G}(A_\alpha^{i*}) - \mathcal{G}(A_\alpha^i)$ 
and $\Delta_{\tau-}\mathcal{G}_{\tau}(A_\alpha) = \mathcal{G}(A_\alpha^{f}) - 
\mathcal{G}(A_\alpha^{f*})$, where we adopt the convention 
$\mathcal{G}(A_\alpha^{i}) \leq \mathcal{G}(A_\alpha^{i*}) < \mathcal{G}(A_\alpha^{f*}) \leq 
\mathcal{G}(A_\alpha^{f})$, which need not be imposed as the optimal solution ensures this 
hierarchical inequality. Slow driving is then followed along the geodesic from 
$\mathcal{G}(A_\alpha^{i*})$ to $\mathcal{G}(A_\alpha^{f*})$ over the time interval $[0^+, 
\tau^-]$. Within the MinAP \cite{atm_2024_var_epr, atm_2025_var_epr_derivation}, the 
thermodynamic EP cost associated with \textit{kinks} is the boundary term 
$\Sigma_{bnd} = \int\mathcal{L}_{bnd}^*[A_\alpha] = \frac{1}{2}[ \Delta_{\tau-}
\mathcal{G}_{\tau}(A_\alpha) ]^2 + \frac{1}{2}[ \Delta_{0+}\mathcal{G}_{\tau}
(A_\alpha)]^2$, which penalizes the formation of \textit{kinks}. The bulk EP required 
to drive along the geodesic from $\mathcal{G}(A_{\alpha}^{i*})$ to $\mathcal{G}(A_{\alpha}^{f*})$ 
is $\Sigma_{bulk} = \int_{0^+}^{\tau^-}\mathcal{L}_{bulk}^* \, dt = \frac{1}{2\tau}
[ \mathcal{G}(A_\alpha^{f*}) - \mathcal{G}(A_{\alpha}^{i*}) ]^2$. Since 
$|\mathcal{G}(A_\alpha^{i*}) - \mathcal{G}(A_\alpha^{f*})| < |\mathcal{G}(A_\alpha^{i}) - 
\mathcal{G}(A_\alpha^{f})|$, it follows that $\Sigma_{bulk} < \Sigma_{qs}^*$. Hence, the 
increase in $\Sigma_{bnd}$ due to \textit{kinks} is compensated by a decrease in $\Sigma_{bulk}$, 
and their interplay plays a key role in finite--time optimal protocols.

The total EP associated with the GFTOC problem is $\Sigma_\tau = \Sigma_{bnd} + \Sigma_{bulk}$. 
Within the MinAP \cite{atm_2024_var_epr, atm_2025_var_epr_derivation}, the minimization of 
$\Sigma_\tau$ requires solving the variational optimal control problem with unknown free 
parameters $A_\alpha^{i*}$ and $A_\alpha^{f*}$,
\begin{equation}\label{eq:finite_time_optimal-control}
\begin{split}
    \Sigma_\tau^* = \inf_{\{A_\alpha^{i*}, A_\alpha^{f*}\}} \left(\Sigma_{\tau}\right).
\end{split}    
\end{equation}
The variation with respect to $A_\alpha^{i*}$ and $A_\alpha^{f*}$ leads to the following set 
of linear Euler--Lagrange equations ${\delta\Sigma_\tau}/{\delta\mathcal{G}(A_\alpha^{i*})} = 0$, 
${\delta\Sigma_\tau}/{\delta\mathcal{G}(A_\alpha^{f*})} = 0$, whose reorganization yields,
\begin{equation}\label{eq:optimization_linear_equations}
\begin{split}
    \left(1 + \frac{1}{\tau}\right)\mathcal{G}(A_\alpha^{i*}) - \frac{1}{\tau}\mathcal{G}(A_\alpha^{f*}) & = \mathcal{G}(A_\alpha^{i}), \\
    -\frac{1}{\tau}\mathcal{G}(A_\alpha^{i*}) + \left(1 + \frac{1}{\tau}\right)\mathcal{G}(A_\alpha^{f*}) & = \mathcal{G}(A_\alpha^{f}).
\end{split}    
\end{equation}
Inverting \cref{eq:optimization_linear_equations}, the solution reads:
\begin{equation}\label{eq:optimal_solution}
\begin{split}
    \mathcal{G}(A_\alpha^{i*}) & = \frac{1+\tau}{2+\tau}\mathcal{G}(A_\alpha^{i}) + \frac{1}{2+\tau}\mathcal{G}(A_\alpha^{f}),
    \\
    \mathcal{G}(A_\alpha^{f*}) & = \frac{1}{2+\tau}\mathcal{G}(A_\alpha^{i}) + \frac{1+\tau}{2+\tau}\mathcal{G}(A_\alpha^{f}).
\end{split}    
\end{equation}
The corresponding finite--time optimal protocol reads:
\begin{equation}\label{eq:optimal_geodesic_finite_time}
\begin{split}
    \mathcal{G}_\tau(A_\alpha) 
    & = \mathcal{G}(A_\alpha^{i}), \hspace{3.4cm} t = 0,
    \\
    & = \left(1 - \frac{t}{\tau}\right)\mathcal{G}(A_\alpha^{i*}) + \frac{t}{\tau}\mathcal{G}(A_\alpha^{f*}), \hspace{0.5cm} t \in (0, \tau),
    \\
    & = \mathcal{G}(A_\alpha^{f}), \hspace{3.4cm} t = \tau.
\end{split}    
\end{equation}
Equivalently, using \cref{eq:optimal_solution}, the finite--time optimal protocol for 
$t \in (0, \tau)$ can be expressed in terms of the known control parameters 
$\mathcal{G}(A_\alpha^{i})$ and $\mathcal{G}(A_\alpha^{f})$ as
\begin{equation}\label{eq:optimal_geodesic_finite_time_explicit}
\begin{split}
    \mathcal{G}_\tau(A_\alpha) 
    & = \left(\frac{1+\tau}{2+\tau} - \frac{t}{2+\tau}\right)\mathcal{G}(A_\alpha^{i})
    +
    \left(\frac{1}{2+\tau} + \frac{t}{2+\tau}\right)\mathcal{G}(A_\alpha^{f}).
\end{split}    
\end{equation}

\Cref{eq:optimal_geodesic_finite_time_explicit} is the optimal finite--time transport map. It 
is equivalent to substituting $t/\tau \to t/(\tau+2)$, $1 \to (1+\tau)/(2+\tau)$, and 
$0 \to 1/(2+\tau)$ in \cref{eq:optimal_protocol}. Physically, optimal finite--time driving is 
equivalent to a total driving time $\tau+2$ with initial and final times $t=1$ and $t=\tau+1$, 
respectively, leading to jumps in the optimal protocol at the endpoints. This reveals the 
structural geometric similarities between finite--time optimal driving and the slow--driving 
process. Using \cref{eq:optimal_geodesic_finite_time_explicit}, the amplitude of \textit{kinks} 
is,
\begin{equation}
\begin{split}
    \Delta_{0+}\mathcal{G}_{\tau}(A_\alpha) = \Delta_{\tau-}\mathcal{G}_{\tau}(A_\alpha) = \frac{1}{2+\tau}\left[\mathcal{G}(A_\alpha^{f}) - \mathcal{G}(A_\alpha^{i})\right].
\end{split}    
\end{equation}
Hence, \textit{kinks} are of equal amplitude in geodesic space, which physically corresponds 
to an equal distribution of the thermodynamic cost associated with \textit{kinks}. Upon 
converting back to the control parameter space $A_\alpha$ using $\mathcal{G}^{-1}$, 
\textit{kinks} have different amplitudes. The higher-mass endpoint has a smaller jump size 
compared to the lower-mass endpoint. The amplitude of \textit{kinks} scales as $1/(2+\tau)$, 
vanishing in the quasi-static limit $\tau \to \infty$. In the fast-driving limit $\tau \to 0$, 
the optimal protocol follows midpoint interpolation in geodesic space,
\begin{equation}\label{eq:optimal_geodesic_short_time}
\begin{split}
    \mathcal{G}_{\tau \to 0^+}(A_\alpha) = \frac{1}{2}\left[\mathcal{G}(A_\alpha^{i}) + \mathcal{G}(A_\alpha^{f})\right].
\end{split}    
\end{equation}
Inverting back to the control parameter space gives $A_\alpha = \mathcal{G}^{-1}\!\left[ \frac{1}{2}\mathcal{G}(A_\alpha^{i}) + \frac{1}{2}\mathcal{G}(A_\alpha^{f})\right]$ 
in the limit $\tau \to 0$. $\mathcal{G}_\tau(A_\alpha)$ is plotted in 
\cref{fig:2}\textcolor{red}{(a)} for different values of $\tau$. Due to 
$\mathcal{G}(A_\alpha) > A_\alpha$, endpoint jumps are amplified for fEQ systems.

Using the finite--time optimal transport map \cref{eq:optimal_geodesic_finite_time_explicit}, 
the optimal bulk and boundary EP are
\begin{equation}\label{eq:optimal_finite_time_bulk_boundary_EP}
\begin{split}
    \Sigma_{bulk}^* & = \frac{\tau}{2(2+\tau)^2}\left[\mathcal{G}(A_\alpha^f) - \mathcal{G}(A_\alpha^i)\right]^2,
    \\
    \Sigma_{bnd}^* & = \frac{2}{2(2+\tau)^2}\left[\mathcal{G}(A_\alpha^f) - \mathcal{G}(A_\alpha^i)\right]^2.
\end{split}    
\end{equation}
\Cref{eq:optimal_finite_time_bulk_boundary_EP} reveals the fundamental tradeoff between bulk 
and boundary EP in finite--time optimal processes. In particular, $\Sigma_{bulk}^* \propto \tau$ 
and $\Sigma_{bnd}^* \propto 1$, implying that finite--time optimal protocols with \textit{kinks} 
are a physical manifestation of the optimization interplay between them. The large-$\tau$ and 
small-$\tau$ regimes correspond to the dominance of $\Sigma_{bulk}^*$ and $\Sigma_{bnd}^*$, 
respectively, in the thermodynamic optimization. The total optimal driving EP is 
$\Sigma_{\tau}^* = \Sigma_{bulk}^* + \Sigma_{bnd}^*$,
\begin{equation}\label{eq:optimal_finite_time_total_EP}
\begin{split}
    \Sigma_{\tau}^* = \frac{1}{2(2+\tau)}\left[\mathcal{G}(A_\alpha^f) - \mathcal{G}(A_\alpha^i)\right]^2.
\end{split}    
\end{equation}
Compared to \cref{eq:optimal_driving_EP}, $\Sigma_{\tau}^*$ is finite in the $\tau \to 0$ 
limit and smaller than $\Sigma_{qs}^*$. They satisfy the non-equilibrium scaling relations 
$\Sigma_{\tau}^* \propto 1/(2+\tau)$ and $\Sigma_{qs}^* \propto 1/\tau$, respectively. 
Importantly, \cref{eq:optimal_finite_time_total_EP} imposes an upper bound on the driving 
dissipation,
\begin{equation}\label{eq:driving_EP_upper_bound}
\begin{split}
    \Sigma_\tau^* \leq \frac{1}{4}\left[\mathcal{G}(A_\alpha^f) - \mathcal{G}(A_\alpha^i)\right]^2,
\end{split}
\end{equation}
which is saturated in the $\tau \to 0$ limit. An analogous upper bound is absent for slow--driving 
processes.

Thermodynamically, finite--time optimal driving is equivalent to slow driving with 
driving time $\tau+2$ instead of $\tau$. This interpretation is consistent with the dynamic 
mapping between slow--driving and finite--time optimal protocols elucidated using 
\cref{eq:optimal_protocol,eq:optimal_geodesic_finite_time_explicit}. The short-time 
underestimation from the $\Sigma_\tau \propto 1/\tau$ scaling has been experimentally observed 
\cite{Ma_2020}. In conclusion, we analytically solve the GFTOC problem by revealing its exact 
geometric connection to its slow--driving counterpart.
\subsection{The physical interpretation of \textit{kinks} and \newline the role of boundary conditions}\label{sec:physical_interpretation}
We elaborate on the physical interpretation of \textit{kinks} within the GFTOC framework. In 
the small driving-time limit, the fundamental principle of ST --- the timescale separation 
between the environment and system degrees of freedom --- is violated. This principle normally 
ensures that the system instantaneously relaxes to the control parameters imposed by the 
environment. Simultaneously, the boundary conditions enforce a constraint of reaching the final 
control parameters within the given finite time $\tau$ from the specified initial control 
parameters. Hence, for small $\tau$, due to violation of the geometric speed limit bound [\cref{eq:slow_driving_speed_limit}], the environment cannot instantaneously impose the required 
control parameters on the system.

\textit{Kinks} resolve this issue by reducing the geodesic distance for the finite--time driving 
process, mathematically expressed as $|\mathcal{G}(A_\alpha^{i*}) - \mathcal{G}(A_\alpha^{f*})| 
< |\mathcal{G}(A_\alpha^{i}) - \mathcal{G}(A_\alpha^{f})|$. This tradeoff between the drive to 
reach the final point along the geodesic and the inability to do so within time $\tau$ produces 
\textit{kinks}. The thermodynamic cost of each \textit{kink} is $\Sigma_{bnd}^*/2$. The system 
undergoes a `thermodynamic shock' at the initial and final points, generating a spontaneous 
discontinuous change in the control parameters, which induces a global transition of the 
system's state. Within our framework, $\Sigma_{bnd}^*$ and $\Sigma_{bulk}^*$ can be interpreted 
analogously to heat and work, respectively: heat corresponds to the instantaneous discrete 
entropy production cost of a jump, and work to the continuous driving of control parameters. To 
our knowledge, this provides a novel thermodynamic understanding of the heat--work dual 
manifestation of the driving EP in finite--time optimally driven far-from-equilibrium processes.
\subsection{Finite--time geometric speed limit}
\begin{figure*}[t!]
\centering
\includegraphics[width=\linewidth]{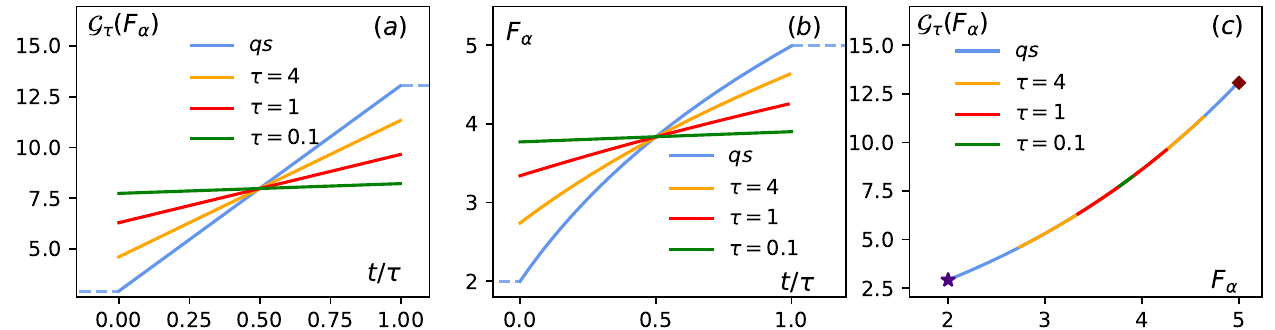}
\caption{(a) The finite--time optimal protocol $\mathcal{G}_{\tau}(F_\alpha) \to t$ is plotted 
for different values of $\tau = \{4, 1, 0.1\}$ with the same initial and final boundary 
conditions (shown by the dotted blue lines), with $F_\alpha^i = 2$ and $F_\alpha^f = 5$. Its 
slow--driving counterpart is denoted by `qs'. As formulated in \cref{sec:housekeeping_epr}, the 
vanishing excess affinity is considered because the optimal control of the housekeeping EPR is 
being solved. \Cref{eq:optimal_protocol,eq:optimal_geodesic_finite_time_explicit,eq:geodesic_expressions} 
are used for the plot, and the driving time scaled by $\tau$ (i.e.\ $t/\tau$) is used on the 
horizontal axis. \textit{Kinks} are of equal amplitude in the geodesic space. (b) The 
corresponding finite--time non-conservative affinity is obtained using the fEQ geodesic 
$\mathcal{G}^{fEQ}(F_\alpha)$ in \cref{eq:geodesic_expressions}, giving the mapping 
$F_\alpha \to t$ in scaled driving time ($t/\tau$). \textit{Kinks} are of unequal amplitude 
in the affinity space; a higher-mass point exhibits a smaller amplitude of \textit{kinks}. 
(c) The corresponding mapping in $\mathcal{G}_{\tau}(F_\alpha) \to F_\alpha$ space; the initial 
and final points are denoted by $\star$ and $\blacklozenge$, respectively. Due to the finite 
driving time constraint, the finite--time optimal protocol \cref{eq:optimal_geodesic_finite_time_explicit} 
traverses a part of the slow--driving geodesic \cref{eq:optimal_protocol} --- a `geometric 
shortcut' --- such that the finite--time geometric speed limit 
\cref{eq:slow_driving_speed_limit_finite_time} is restored, analogously to the slow--driving 
geometric speed limit \cref{eq:slow_driving_speed_limit}. The discontinuous jumps in the 
finite--time driving protocol are further constrained by equally distributing the thermodynamic 
cost [$\Sigma_{bnd}^*$ in \cref{eq:optimal_finite_time_bulk_boundary_EP}] between the initial 
and final endpoints, generating a `thermodynamic shock' at the initial and final times.}
\label{fig:2}
\end{figure*}
Using $\Sigma_{bulk}^*$ from \cref{eq:optimal_finite_time_bulk_boundary_EP} and 
\cref{eq:slow_driving_speed_limit,eq:optimal_solution}, one finds
\begin{equation}\label{eq:slow_driving_speed_limit_finite_time}
\begin{split}
    \tau_{\tau}^* = \frac{\left[ \mathcal{G}(A_\alpha^{f*}) - \mathcal{G}(A_{\alpha}^{i*})\right]^2}{2\Sigma_{bulk}^*} = \tau.
\end{split}    
\end{equation}
This shows that the selection of endpoint jump values $\mathcal{G}(A_\alpha^{i*})$ and 
$\mathcal{G}(A_\alpha^{f*})$ is constrained to restore the geometric speed limit bound 
\cref{eq:slow_driving_speed_limit} for the finite--time optimal driving process. Physically, 
this implies that the allowed endpoint jumps are minimal, just sufficient to restore the 
finite--time geometric speed limit constraint for any given $\tau$. This highlights a 
fundamental symmetry: \textit{kinks} are a non-equilibrium mechanism that circumvents the 
dynamical constraint of limited driving time quantified by the geometric speed limit, 
representing the dynamical counterpart of a `thermodynamic shock'.

The upper bound on the total amplitude of \textit{kinks}, 
$|\Delta_{0^+}\mathcal{G}_\tau(A_\alpha)| + |\Delta_{\tau^-}\mathcal{G}_\tau(A_\alpha)|$, a 
boundary property, can be obtained from the thermodynamic cost of driving and the distance 
covered in the control parameter space (a bulk property). The equal thermodynamic cost 
$\Sigma_{bnd}^*$ for each \textit{kink} results from minimizing $\Sigma_{bnd}^*$ for a 
constrained total amplitude of \textit{kinks}. Hence, \textit{kinks} arise from finite 
driving-time constraints that may prevent reaching the final state along the geodesic. The 
finite--time optimal protocol restores \cref{eq:slow_driving_speed_limit} while minimizing the 
thermodynamic cost of both \textit{kinks}. Importantly, the slow--driving assumption is broken 
for any finite $\tau$ and does not require $\tau < \tau_{qs}^*$. A continuous change of $\tau$ 
therefore induces a smooth transition from 
\cref{eq:optimal_protocol,eq:optimal_driving_EP} to 
\cref{eq:optimal_geodesic_finite_time,eq:optimal_finite_time_total_EP}, with $\tau$ as the 
relevant control parameter. In conclusion, this mechanism is to be interpreted as a `geometric 
shortcut'. \Cref{fig:2} pictorially summarizes the underlying geometric structure of GFTOC. 
\subsection{Key features of geodesic}
\textit{Time-reversal symmetry}. \textemdash \:
\Cref{eq:optimal_protocol,eq:optimal_geodesic_finite_time_explicit} satisfy time-reversal 
symmetry, as they are invariant under the transformation $\mathcal{G}(A_\alpha^i) \to 
\mathcal{G}(A_\alpha^f)$, $\mathcal{G}(A_\alpha^f) \to \mathcal{G}(A_\alpha^i)$, and 
$t \to \tau - t$. This symmetry arises fundamentally from the geodesic structure of the 
slow--driving process. The symmetric amplitude of \textit{kinks} extends this symmetry to the 
finite--time geodesic \cref{eq:optimal_geodesic_finite_time}.

\textit{Experimental inference and non-equilibrium scaling relation}.\textemdash
The GFTOC framework implies that the minimum driving EP is obtained by computing the geodesic. 
If experimental setups can measure the optimal driving EP (thermodynamically constrained to 
support a minimum dissipation for the driving process), one can pose the inverse problem of 
reconstructing the geodesic for given initial and final control parameters. The non-equilibrium 
scaling relation \cref{eq:optimal_finite_time_total_EP} can be used to infer the analytical 
shape of $\mathcal{G}(A_\alpha)$. For an observed set of data $\{(\Sigma_{\tau}^*, \tau)\}$ at 
fixed $A_\alpha^i$ and $A_\alpha^f$, the collapse of the data yields 
$\mathcal{G}(A_\alpha^f) - \mathcal{G}(A_\alpha^i) = \sqrt{2(2+\tau)\Sigma_{\tau}^*}$. 
Repeating this procedure for different $A_\alpha^i$ and $A_\alpha^f$ reconstructs 
$\mathcal{G}(A_\alpha)$ experimentally or numerically. Moreover, at fixed $A_\alpha^i$ and $A_\alpha^f$, computing $\Sigma_{\tau\to 0}^*$ 
via \cref{eq:driving_EP_upper_bound} offers the practical advantage of measuring the geodesic of control parameters from experimental data $\{\Sigma_{\tau\to 0}^*\}$ of a swift change of control parameters, without requiring slow--driving calibrations of the geodesic.
The geodesic shape informs about the 
non-equilibrium character of the process and the contribution of the EP, since it has different 
analytical dependence on the cEQ and fEQ regimes, as discussed in 
\cref{sec:geodesic_expressions} and further in \cref{sec:application_GFTOC}.

{
\textit{Experimental implementation of kinks}.\textemdash The rapid change of control 
parameters and the generation of \textit{kinks} in infinitesimal time poses an experimental 
restriction due to the infinite power required, but this is not a limitation of the GFTOC 
framework, since the `thermodynamic shock' [$\Sigma_{bnd}^*$, \cref{eq:optimal_finite_time_bulk_boundary_EP}] 
(power multiplied by infinitesimal time) is always finite. However, more realistic experimental 
setups operate on a discrete-time basis with an intrinsic finite timescale $\Delta t$ of the 
experimental clock, which implies that the power required to implement \textit{kinks} is large 
but finite; \textit{kinks} are therefore experimentally realisable in such a setup.
}

{
\textit{Experimental verification of GFTOC}.\textemdash We highlight a simple test for the 
experimental verification of GFTOC in the context of Ref.~\cite{Ma_2020}. As shown by 
\cref{eq:optimal_finite_time_total_EP}, the finite-time scaling of the optimal driving EP is modified 
from $1/\tau$ to $1/(\tau+2)$, which is verifiable using insets (b) and (c) of Fig.~2 in 
Ref.~\cite{Ma_2020}. 
}
\subsection{Comparison and differences to model-specific known results}\label{sec:comparison_papers}
finite--time optimal protocols with \textit{kinks} were analytically computed for a particle in 
a harmonic trap \cite{Schmiedl_2007} and recently unified using optimal transport theory 
\cite{Zhong_2024}. These model-specific results differ from the generalized finite--time 
framework in three aspects: (i) they optimize the free energy $\psi_E$ using control parameters 
of the equilibrium energy functional $E$, a close-to-equilibrium methodology; (ii) the boundary 
conditions used to obtain \textit{kinks} differ; and (iii) the thermodynamic cost of 
\textit{kinks} is undefined in Refs.~\cite{Schmiedl_2007, Zhong_2024}, leading to the 
physical interpretation of \textit{kinks} as a boundary artefact.

We briefly summarize the procedure outlined in Ref.~\cite{Zhong_2024}. The boundary conditions 
used in Refs.~\cite{Schmiedl_2007, Zhong_2024} are equivalent to fixing $A_\alpha^{i*} = A_i$ 
and computing $A_\alpha^{*int} \in (A_\alpha^i, A_\alpha^f)$ that minimizes the bulk driving EP 
under the final value constraint. Thus, $A_\alpha^i \to A_\alpha^{*int}$ is termed the feasible 
driving protocol for the given finite time $\tau$. Then, $A_\alpha^{*int}$ is used to compute 
$A_\alpha^{*i}$ and $A_\alpha^{*f}$ by introducing counter-diabatic driving terms that shift 
the initially computed driving protocol $A_\alpha^i \to A_\alpha^{*int}$ to 
$A_\alpha^{i*} \to A_\alpha^{f*}$ in driving time $\tau$, generating \textit{kinks} of 
amplitude $A_\alpha^i - A_\alpha^{*i}$ and $A_\alpha^{f} - A_\alpha^{f*}$ in the optimal 
protocol. This implies that the computation of the initial geodesic for the finite--time process 
is based on the assumption that \textit{kinks} are absent at the initial time $t = 0$. However, 
the finite--time optimal protocol $A_\alpha^{i*} \to A_\alpha^{f*}$ is obtained by shifting this 
protocol via counter-diabatic driving to obtain \textit{kinks} at both $t = 0^+$ and 
$t = \tau^-$.

This contradicts the formulation of constrained and unconstrained optimization for the initial 
($t = 0^+$) and final ($t = \tau^-$) points, respectively. Hence, mathematically, this is 
interpreted as a boundary condition artefact arising from the inability to assign a 
thermodynamic cost to \textit{kinks} within their framework. In contrast, within the MinAP 
formalism, a thermodynamic cost is inherently assigned to \textit{kinks}, and unconstrained 
optimization is implemented for both the initial ($t = 0^+$) and final ($t = \tau^-$) points, 
as signified by the double optimization problem in \cref{eq:finite_time_optimal-control}. 
Therefore, our optimization framework uses minimal assumptions, equivalently assuming the least 
information about the optimal protocols. This allows us to implement an unconstrained 
optimization problem in \cref{eq:finite_time_optimal-control} that searches a wider control 
parameter space for the optimal protocol.

These differences lead to rather drastic physical consequences. For example, the short-time 
optimal protocol \cref{eq:optimal_geodesic_short_time} is a midpoint interpolation in the 
geodesic space $\mathcal{G}(A_\alpha)$ and not necessarily in the control parameter space 
$A_\alpha$, unless $\mathcal{G}(A_\alpha) = A_\alpha$. Hence, even for short driving time 
$\tau$, the finite--time optimal protocol \cref{eq:optimal_geodesic_short_time} respects the 
Riemannian geometric structure in the control parameter space. In contrast, the short-time 
protocols obtained in Refs.~\cite{Schmiedl_2007, Zhong_2024} follow a midpoint interpolation 
in the control parameter space. This difference is irrelevant in the case of the optimal 
control of the trap centre, since $\mathcal{G}(A_\alpha) = A_\alpha$ for the control of the 
trap centre and the short-time optimal protocols obtained in 
Refs.~\cite{Blaber_2021, Blaber_2023}. However, $\mathcal{G}(A_\alpha) \neq A_\alpha$ for the 
optimal control of the trap stiffness 
\cite{Sekimoto_1997_slow_driving_optimal_control, Sivak_2012}, giving divergent analytical 
results compared to Refs.~\cite{Schmiedl_2007, Zhong_2024, Blaber_2021, Blaber_2023}. 
Subsequently, we revisit this issue in \cref{sec:harmonic_trap} with the example of a 
stochastic particle in a harmonic trap and reconcile the discrepancies between optimal transport 
theory \cite{Dechant_2019, Nakazato_2021_omtp_st, Oikawa_2025} and thermodynamic geometry 
\cite{Sekimoto_1997_slow_driving_optimal_control, Sivak_2012}.
\section{Generalizations} \label{sec:optimal_control_generalization}
\subsection{ Multiple control parameters }\label{sec:optimal_control_multiparameter}
The driving EPR in \cref{eq:optimal_driving_epr} assumes that all control parameters are 
decoupled and independently driven. However, in practice, there may be cross-couplings between 
different control parameters. In addition, the transition affinity $A_\alpha$ accounts only for 
the anti-symmetric part in the transition space \cite{ATM_2024_nr_st}, and the transition 
mobility $D_\alpha$ was fixed. We aim to relax this constraint.

For slow driving of multiple parameters, the driving Lagrangian is given by
\begin{equation}\label{eq:optimal_driving_epr_multi}
\begin{split}
    \mathcal{L}_{drv}^*\left[\vec{\lambda}\right] = \frac{1}{2} g_{ij} \dot{\lambda}^{i} \dot{\lambda}^{j},
\end{split}    
\end{equation}
where \cref{eq:optimal_driving_epr_multi} is written in Einstein summation convention, which 
signifies summation over contracted indices. $\vec{\lambda}$ is the vector of control parameters 
with $i$-th component $\lambda^i$. The Hessian of $\mathcal{L}^*$ (the mass) gives the metric 
tensor $g_{ij}$,
\begin{equation}\label{eq:metric_tensor}
\begin{split}
    g_{ij} 
    & =
    \begin{bmatrix}
        \partial_{\lambda^i}^2 \mathcal{L}^* 
        & \partial_{\lambda^j}\partial_{\lambda^i} \mathcal{L}^*
        \\
        \partial_{\lambda^j} \partial_{\lambda^i} \mathcal{L}^* 
        & \partial_{\lambda^j}^2 \mathcal{L}^*
    \end{bmatrix}.
\end{split}    
\end{equation}
The MinAP implies solving a variational optimization problem, $\Sigma_{qs}^* = 
\inf(\Sigma_{drv}^{qs}) = \int_0^\tau \mathcal{L}_{drv}^* \, dt$. The geodesic equation for 
the control parameter that minimizes $\Sigma_{drv}^{qs}$ in \cref{eq:optimal_driving_epr_multi} 
reads,
\begin{equation}\label{eq:geodesic_equation_multi}
\begin{split}
    \ddot{\lambda}^i + \Gamma_{j k}^{i} \dot{\lambda}^j \dot{\lambda}^k = 0,
\end{split}    
\end{equation}
where the Christoffel symbols are defined as
\begin{equation}\label{eq:christoffel_symbol}
\begin{split}
    \Gamma_{j k}^{i} = \frac{1}{2} g^{im} \left(\frac{\partial g_{mj}}{\partial{\lambda}^k} + \frac{\partial g_{mk}}{\partial \lambda^j} - \frac{\partial g_{jk}}{\partial \lambda^m}\right).
\end{split}    
\end{equation}
\Cref{eq:optimal_driving_epr_multi,eq:geodesic_equation_multi} are equivalent to 
\cref{eq:optimal_driving_epr,eq:geodesic_equation} for the multi--parameter slow--driving optimal 
control problem.

By construction, the geodesic $\mathcal{G}(\vec{\lambda})$ is the solution that minimizes 
$\Sigma_{drv}^{qs}$, $\mathcal{G}(\vec{\lambda}) = \arg\inf_{\{\lambda\}}\left(\Sigma_{drv}^{qs}\right)$. 
The corresponding optimal protocol is a linear interpolation, which reads,
\begin{equation}\label{eq:optimal_protocol_multi_parameter}
\begin{split}
    \mathcal{G}(\vec{\lambda}) 
    = \left(1 - \frac{t}{\tau}\right)\mathcal{G}(\vec{\lambda}^{inl}) + \frac{t}{\tau}\mathcal{G}(\vec{\lambda}^{fnl}).
\end{split}    
\end{equation}
To avoid confusion with the vector index, we denote the vectors of initial and final control 
parameters by $\vec{\lambda}^{inl}$ and $\vec{\lambda}^{fnl}$, respectively. Using 
\cref{eq:optimal_protocol_multi_parameter}, the optimal slow--driving EP reads:
\begin{equation}\label{eq:optimal_driving_EP_multi_parameter}
\begin{split}
    \Sigma_{qs}^* = \frac{1}{2\tau}\left|\left|\mathcal{G}(\vec{\lambda}^{fnl}) - \mathcal{G}(\vec{\lambda}^{inl})\right|\right|_2^2.
\end{split}    
\end{equation}
Here, $||\cdot||_2$ is the $L^2$-norm that quantifies the distance on the Riemannian manifold 
induced by multiple control parameters. The finite--time optimal control framework developed in 
\cref{sec:finite_time_optimal_control} is based on the MinAP combined with the existence of the 
slow--driving geodesic $\mathcal{G}(\vec{\lambda})$ and incorporating the possibility of 
\textit{kinks}. Hence, a multi--parameter finite--time optimal control protocol is rather 
straightforward, provided that a slow--driving geodesic \cref{eq:optimal_protocol_multi_parameter} 
exists and is computed. The corresponding finite--time multi--parameter optimal protocol reads,
\begin{equation}\label{eq:optimal_geodescic_finite_time_multi_parameter}
\begin{split}
    \mathcal{G}_\tau(\vec{\lambda}) 
    & = \left(\frac{1+\tau}{2+\tau} - \frac{t}{2+\tau}\right)\mathcal{G}(\vec{\lambda}^{inl}) + \left(\frac{1}{2+\tau} + \frac{t}{2+\tau}\right)\mathcal{G}(\vec{\lambda}^{fnl}),
\end{split}    
\end{equation}
with the finite--time multi--parameter optimal driving EP:
\begin{equation}\label{eq:optimal_driving_finite_time_EP_multi_parameter}
\begin{split}
    \Sigma_{\tau}^* = \frac{1}{2(2+\tau)}\left|\left|\mathcal{G}(\vec{\lambda}^{fnl}) - \mathcal{G}(\vec{\lambda}^{inl})\right|\right|_2^2.
\end{split}    
\end{equation}
\Cref{eq:optimal_protocol_multi_parameter,eq:optimal_driving_EP_multi_parameter,eq:optimal_geodescic_finite_time_multi_parameter,eq:optimal_driving_finite_time_EP_multi_parameter} 
generalize \cref{eq:optimal_protocol,eq:optimal_driving_EP,eq:optimal_geodesic_finite_time_explicit,eq:optimal_finite_time_total_EP}, 
respectively, to the multi--parameter generalized optimal control problem.
\subsection{Numerical computation of geodesic}
We have focused on the exact analytical formulation of the GFTOC framework. However, 
analytically computing the exact geodesic $\mathcal{G}(\vec{\lambda})$ is not always feasible 
for more sophisticated problems. To this end, a numerical framework is required to compute the 
geodesic and optimal protocols. Here, we highlight two different numerical algorithms 
implemented to compute the geodesic. \Cref{alg:algorithm_action_minimization_optimal_protocols} 
is based on the MinAP, focusing on the thermodynamic cost of driving, where the geodesic 
$\mathcal{G}(\vec{\lambda})$ is obtained by minimizing the Lagrangian 
\cref{eq:optimal_driving_epr_multi} and extended to return the finite--time optimal protocol 
$\mathcal{G}_\tau(\vec{\lambda})$; it reads,
\begin{algorithm}[H] 
\caption{Computing finite--time optimal protocols by minimizing the cost function (action) \\ 
  \textbf{Input:} $\vec{\lambda}^{inl}, \vec{\lambda}^{fnl}, \tau, g_{ij}$, \cref{eq:optimal_driving_epr_multi,eq:optimal_geodescic_finite_time_multi_parameter,eq:optimal_driving_finite_time_EP_multi_parameter}}
\begin{algorithmic}[1]
   \State Compute the geodesic $\mathcal{G}(\vec{\lambda})$ by recursive minimization of the 
   action \cref{eq:optimal_driving_epr_multi}, connecting $\mathcal{G}(\vec{\lambda}^{inl})$ 
   and $\mathcal{G}(\vec{\lambda}^{fnl})$.
   \State Calculate the finite--time optimal protocol $\mathcal{G}_\tau(\vec{\lambda})$ and EP 
   cost $\Sigma_\tau^*$ using \cref{eq:optimal_geodescic_finite_time_multi_parameter,eq:optimal_driving_finite_time_EP_multi_parameter}, respectively.
   \State Return the optimal protocol $\mathcal{G}_\tau(\vec{\lambda})$ and EP cost $\Sigma_\tau^*$.
\end{algorithmic}
\label{alg:algorithm_action_minimization_optimal_protocols}
\end{algorithm}
%
%
{
\Cref{alg:algorithm_action_minimization_optimal_protocols} extends the class of numerical action 
optimization methods developed in Refs.~\cite{Jordan_1998, Benamou_2000, Weinan_2004, 
Heymann_2008, Vanden-Eijnden_2008, Rotskoff_2017, Grafke_2017, Grafke_2019, Gagrani_2023, 
Zakine_2023, Smith_2024} and using machine learning techniques in Refs.~\cite{Cherukuri_2017, 
Lin_2020, Yan_2022}, to compute finite--time optimal protocols for any generic model. Precisely, 
the recursive minimization of the action implemented in these algorithms corresponds to 
\textit{Step~1} in \cref{alg:algorithm_action_minimization_optimal_protocols}. However, none of 
the previous numerical algorithms incorporate \textit{kinks}, as a result of their failure to 
assign a thermodynamic cost to \textit{kinks}; their applicability is therefore restricted to 
the computation of slow--driving optimal protocols. In contrast, 
\cref{alg:algorithm_action_minimization_optimal_protocols} computes finite--time optimal protocols 
by exploiting the exact analytical mapping between slow--driving and finite--time optimal control 
developed by the GFTOC framework [\cref{eq:optimal_geodescic_finite_time_multi_parameter}]. 
Despite the common motif of MinAP for the algorithms in Refs.~\cite{Jordan_1998, Benamou_2000, 
Weinan_2004, Heymann_2008, Vanden-Eijnden_2008, Rotskoff_2017, Grafke_2017, Grafke_2019, 
Gagrani_2023, Zakine_2023, Smith_2024, Cherukuri_2017, Lin_2020, Yan_2022}, the efficient 
implementation of \textit{Step~1} has some technical differences that are not relevant here.
}
To our knowledge, \cref{alg:algorithm_action_minimization_optimal_protocols} is the first 
formulation that numerically computes finite--time optimal protocols with \textit{kinks} and the 
associated finite--time driving EP cost.

{
Note that \cref{alg:algorithm_action_minimization_optimal_protocols} can be modified to 
explicitly compute $\mathcal{G}_\tau(\vec{\lambda})$ instead of $\mathcal{G}(\vec{\lambda})$ 
[without using the analytical result of 
\cref{eq:optimal_geodescic_finite_time_multi_parameter}]. To this end, the thermodynamic cost 
of \textit{kinks} is assigned under the action minimization problem, and the endpoint 
discontinuity in the protocol space is allowed by relaxing the boundary condition constraint in 
\cref{eq:optimal_driving_epr_multi}. Precisely, this requires allowing \textit{kinks} in the 
time-discretization of $\vec{\lambda}$ (by inserting two extra time points $\vec{\lambda}(0^+)$ 
and $\vec{\lambda}(\tau^-)$) for the numerical implementation, and accounting for their 
thermodynamic cost through \cref{eq:optimal_driving_epr_multi}; a modification equivalent to 
implementing \textit{Step~2} and \textit{Step~3} of 
\cref{alg:algorithm_action_minimization_optimal_protocols} inside the recursive optimization 
loop (\textit{Step~1}). The recursive minimization of the action using the algorithms in 
Refs.~\cite{Jordan_1998, Benamou_2000, Weinan_2004, Heymann_2008, Vanden-Eijnden_2008, 
Rotskoff_2017, Grafke_2017, Grafke_2019, Gagrani_2023, Zakine_2023, Smith_2024, Cherukuri_2017, 
Lin_2020, Yan_2022} (\textit{Step~1}) then leads to the finite--time geodesic 
$\mathcal{G}_\tau(\vec{\lambda})$ that exhibits \textit{kinks}.
}

\Cref{alg:algorithm} computes finite--time optimal protocols by numerically solving the geodesic 
equation \cref{eq:geodesic_equation_multi} as a boundary value problem to obtain 
\cref{eq:optimal_protocol_multi_parameter}, focusing on the driving dynamics. The algorithm 
reads:
\begin{algorithm}[H] 
\caption{Computing finite--time optimal protocols by numerically solving geodesic ODEs\\ 
  \textbf{Input:} $\vec{\lambda}^{inl}, \vec{\lambda}^{fnl}, \tau, g_{ij}, \Gamma_{ij}^k$, \cref{eq:geodesic_equation_multi} and \cref{eq:optimal_geodescic_finite_time_multi_parameter}}
\begin{algorithmic}[1]
   \State Compute the geodesic $\mathcal{G}(\vec{\lambda})$ by numerically solving 
   \cref{eq:geodesic_equation_multi} under the boundary value constraints $\vec{\lambda}^{inl}$ 
   and $\vec{\lambda}^{fnl}$ at $t = 0$ and $t = \tau$.
   \State Calculate the finite--time optimal protocol $\mathcal{G}_\tau(\vec{\lambda})$ and EP 
   cost $\Sigma_\tau^*$ using \cref{eq:optimal_geodescic_finite_time_multi_parameter,eq:optimal_driving_finite_time_EP_multi_parameter}, respectively.
   \State Return the optimal protocol $\mathcal{G}_\tau(\vec{\lambda})$ and EP cost $\Sigma_\tau^*$.
\end{algorithmic}
\label{alg:algorithm}
\end{algorithm}
In contrast to \cref{alg:algorithm_action_minimization_optimal_protocols}, 
\cref{alg:algorithm} can only compute $\mathcal{G}(\vec{\lambda})$ and requires the analytical 
result of \cref{eq:optimal_geodescic_finite_time_multi_parameter} to compute 
$\mathcal{G}_\tau(\vec{\lambda})$.
{
The numerical approach presented in this section broadens the applicability of the GFTOC 
framework to finite--time driven processes when an exact analytical solution for the geodesic is 
unavailable, by integrating the theoretical developments (the exact mapping between slow--driving 
and finite--time driving) with existing numerical optimization methods available for computing 
slow--driving geodesics.
}
{
\subsection{Applicability of the GFTOC framework beyond ST}
The definition of $\mathcal{L}^*$ as the mean EPR $\langle\dot{\Sigma}\rangle$ allows for a 
thermodynamic interpretation, which is the focus of ST and requires the LDB condition. However, 
MinAP is valid for fEQ systems beyond ST \cite{atm_2024_var_epr, atm_2025_var_epr_derivation}. 
Analogously, the GFTOC framework is generic and broadly applicable beyond ST, where 
$\mathcal{L}^*$ need not have a thermodynamic meaning. The optimal control of any relevant 
physical (or unphysical) quantity $\mathcal{L}^*$ in fEQ systems, parameterized by control 
parameters $\vec{\lambda}$, can be explored using GFTOC. Geometric optimization on the 
Riemannian manifold ($\mathcal{L}^*$) of the control parameters using GFTOC then follows 
trivially. Moreover, a more generic cost penalty ($Cost(\phi(t_i), \vec{\lambda})$) at any 
intermediate time $t_i \in [0, \tau]$ can be incorporated into GFTOC, resulting in the 
following modified optimization problem: $\inf_{\vec{\lambda}}\left[Cost(\phi(t_i), \vec{\lambda}) + 
\int_{0}^{\tau}\mathcal{L}_{drv}^{*}\,dt\right]$. For example, $D_E^{KL}$ is a cost penalty 
for imposing the state--space constraint of a targeted Boltzmann distribution $\{\rho_i^E\}$, 
predominantly used in optimal transport theory, with a physical (thermodynamic) connection to 
the excess EP. In summary, GFTOC extends optimal control methodologies beyond their relevance 
in ST, as discussed here.
}
\section{Applications of GFTOC framework}\label{sec:application_GFTOC} \
\subsection{Optimal control of free energy}\label{sec:free_energy}
The EP defined for the driving of the free energy $\psi_E$ by changing the control parameters 
$\{\lambda_E\}$ of $E$ reads, $\mathcal{L}_{drv}^* = \frac{1}{2}\left(\partial^2\psi_E/
\partial\lambda_E^i\,\partial\lambda_E^j\right)\dot{\lambda}_E^i\dot{\lambda}_E^j$. The initial 
and final control parameter vectors are denoted by $\vec{\lambda}_E^{inl}$ and 
$\vec{\lambda}_E^{fnl}$, respectively. The corresponding optimal slow--driving EP and optimal 
driving protocol read:
\begin{equation}\label{eq:slow_driving_free_energy}
\begin{split}
    &\Sigma_{qs}^* = \frac{1}{2\tau}\left[ \mathcal{G}(\vec{\lambda}_E^{inl}) - \mathcal{G}(\vec{\lambda}_E^{fnl})\right]^2, 
    \\
    &\mathcal{G}(\vec{\lambda}_E) 
    = \left(1 - \frac{t}{\tau}\right)\mathcal{G}(\vec{\lambda}_E^{inl})
    +
    \frac{t}{\tau}\mathcal{G}(\vec{\lambda}_E^{fnl}).
\end{split}    
\end{equation}
The corresponding finite--time optimal EP and finite--time optimal driving protocol read:
\begin{equation}\label{eq:finite_time_driving_free_energy}
\begin{split}
    \Sigma_{\tau}^* & = \frac{1}{2(2+\tau)}\left[\mathcal{G}(\vec{\lambda}_E^{inl}) - \mathcal{G}(\vec{\lambda}_E^{fnl})\right]^2, 
    \\
    \mathcal{G}_\tau(\vec{\lambda}_E) 
    & = 
    \left(\frac{1+\tau}{2+\tau} - \frac{t}{2+\tau}\right)\mathcal{G}(\vec{\lambda}_E^{inl})
    +
    \left(\frac{1}{2+\tau} + \frac{t}{2+\tau}\right)\mathcal{G}(\vec{\lambda}_E^{fnl}).
\end{split}    
\end{equation}
Note that the exact geodesic functions $\mathcal{G}(\vec{\lambda}_E)$ are model-specific, due 
to the different analytical dependence of $\psi_E$ on $\lambda_E$ \cite{Salamon_1985, 
Brody_1995, Schlogl_1985, Crooks_2007}. In the next section, we explicitly compute the geodesic 
functions $\mathcal{G}(\vec{\lambda}_E)$ for a stochastic particle in a harmonic trap.
{$\Sigma_{qs}^*$ and $\Sigma_{\tau}^*$ are the `\textit{kinetic energetic}' finite--time dissipation 
corrections to the unavoidable quasi-static (`\textit{potential energetic}') free energy change 
$-\Delta_{\lambda_E} \psi_E = \psi(\vec{\lambda}_E^{inl}) - \psi(\vec{\lambda}_E^{fnl})$.}
\subsection{Stochastic particle in a harmonic trap and \\ Wasserstein distance for the Gaussian distribution}\label{sec:harmonic_trap}
%
%
{
To demonstrate the broader applicability of GFTOC to continuous state--space systems, we consider 
a stochastic particle in a harmonic trap potential, which models an important equivalence class 
describing a large class of continuous state--space physical systems, and has therefore served as 
a testbed for novel theoretical developments in ST.
}
The stochastic particle satisfies the Boltzmann distribution $\rho^E(x) \propto e^{-k(x-m)^2}$, with 
trap stiffness $k$ and centre $m$. These two control parameters fully capture the system's 
statistical properties. We consider a finite--time optimal control problem for the time-dependent 
change of the trap stiffness $k(t)$ and centre $m(t)$ 
\cite{Schmiedl_2007, Sekimoto_1997_slow_driving_optimal_control, Sivak_2012}. The inverse of 
the trap stiffness quantifies the covariance ($C$) of the Gaussian distribution for the particle 
position, $C \propto 1/k$ \cite{Schmiedl_2007}. Hence, by definition, the finite--time optimal 
control formulation for changing the centre and stiffness of the harmonic trap is equivalent to 
the optimal transport problem of changing the mean and covariance of the Gaussian distribution 
for the particle position \cite{Dechant_2019}, or to the application of thermodynamic geometry 
to the harmonic trap \cite{Sekimoto_1997_slow_driving_optimal_control, Sivak_2012},
{provided that the stochastic particle satisfies the Boltzmann distribution $\rho^E(x)$ imposed by the instantaneous control parameters $k(t)$ and $m(t)$ at all times.} However, 
the results obtained from different methods do not agree 
\cite{Schmiedl_2007, Sekimoto_1997_slow_driving_optimal_control, Sivak_2012, Dechant_2019, 
Oikawa_2025}. We resolve these discrepancies using the GFTOC framework.

We consider an optimal control problem for changing the trap stiffness and centre from 
$\{m_i, k_i\}$ to $\{m_f, k_f\}$. The \textit{mass} for the centre and stiffness of the trap are 
$g_{mm} = 1$ and $g_{kk} = 1/k^3$ \cite{Sekimoto_1997_slow_driving_optimal_control, Sivak_2012}, 
implying the geodesic functions $\mathcal{G}(m) = m$ and $\mathcal{G}(k) = k^{-1/2}$. Using 
$C \propto 1/k$ due to the instantaneous relaxation, we recover the geodesic function $\mathcal{G}(C) = C^{1/2}$ for the 
covariance. Hence, the minimum slow--driving EP and the corresponding optimal protocols for the 
mean and covariance read:
\begin{subequations}
\begin{equation}\label{eq:slow_driving_wasserstein_distance}
    \Sigma_{qs}^* = \frac{1}{2\tau}\left[(m_f - m_i)^2 + \left(\sqrt{C_f} - \sqrt{C_i}\right)^2\right],
\end{equation}
\begin{equation}\label{eq:slow_driving_transport_map_center}
    m(t) = \left(1 - \frac{t}{\tau}\right)m_i + \frac{t}{\tau}m_f,    
\end{equation}
\begin{equation}\label{eq:slow_driving_transport_map_stiffness}
    \sqrt{C(t)} = \left(1 - \frac{t}{\tau}\right)\sqrt{C_i} + \frac{t}{\tau}\sqrt{C_f}.
\end{equation}
\end{subequations}
\Cref{eq:slow_driving_wasserstein_distance} is the square of the $\mathcal{W}_2$ Wasserstein 
distance between the initial and final Gaussian probability distributions, divided by $2\tau$, 
as in Refs.~\cite{Dechant_2019, Nakazato_2021_omtp_st, Ito_2024_omtp_st}, with the 
corresponding optimal protocols for the mean and covariance given by 
\cref{eq:slow_driving_transport_map_center,eq:slow_driving_transport_map_stiffness}, as in 
Ref.~\cite{Dechant_2019}. 

Applying the GFTOC framework, the minimum finite--time driving EP and finite--time optimal 
protocols for the mean and covariance read:
\begin{subequations}
\begin{equation}\label{eq:finite_time_wasserstein_distance}
    \Sigma_{\tau}^* = \frac{1}{2(2+\tau)}\left[(m_f - m_i)^2 + \left(\sqrt{C_f} - \sqrt{C_i}\right)^2\right],
\end{equation}
\begin{equation}\label{eq:finite_time_transport_map_center}
    m_\tau(t) = \left(\frac{1+\tau}{2+\tau} - \frac{t}{2+\tau}\right)m_i + \left(\frac{1}{2+\tau} + \frac{t}{2+\tau}\right)m_f,    
\end{equation}
\begin{equation}\label{eq:finite_time_transport_map_stiffness}
    \sqrt{C_\tau(t)} = \left(\frac{1+\tau}{2+\tau} - \frac{t}{2+\tau}\right)\sqrt{C_i} + \left(\frac{1}{2+\tau} + \frac{t}{2+\tau}\right)\sqrt{C_f}.
\end{equation}
\end{subequations}
\Cref{eq:finite_time_wasserstein_distance} is the novel formulation for finite--time processes, 
requiring the square of the $\mathcal{W}_2$ Wasserstein distance between the initial and final 
Gaussian probability distributions to be divided by $2(2+\tau)$; this is less than 
\cref{eq:slow_driving_wasserstein_distance} due to \textit{kinks}. The finite--time optimal 
protocol for the trap centre \cref{eq:finite_time_transport_map_center} agrees with 
Refs.~\cite{Schmiedl_2007, Zhong_2024}. However, the finite--time optimal protocol for the trap 
stiffness \cref{eq:finite_time_transport_map_stiffness} differs from 
Refs.~\cite{Schmiedl_2007, Zhong_2024} and is novel. This difference in boundary conditions and their 
physical implications have been detailed in \cref{sec:comparison_papers}. 
\subsection{Optimal control of excess EP}\label{sec:excess_epr}
The optimal control formulation has been developed for the affinity, which lies in the 
transition space. However, as shown in \cref{eq:mean_EPR}, a part of the EPR can be integrated 
to obtain the excess EP $\langle \Sigma^{ex} \rangle = -D_E^{KL}$, defined in the state space. The integration 
is implemented: (1) from the transition space to the state space, and (2) in time. This implies 
that $\langle \Sigma^{ex} \rangle$ is the boundary term that does not require knowledge of the transition--space 
topology. This property of the excess EP has been exploited in the literature to formulate the 
control or geodesic description of cEQ systems 
\cite{Rao_1945,Qian_2002, Crooks_2007, Ito_2018, Ito_2024_omtp_st, Melo_2025_FI, Melo_2025_SFI, 
Melo_2025_FR_geometry},
{which is particularly relevant when the state--space distribution does not instantaneously relax 
to the Boltzmann distribution $\{\rho_i^E\}$ imposed by $E$, in which case explicit state--space 
control of the instantaneous distribution $\{\rho_i\}$ becomes important.} For $-\langle \Sigma^{ex} \rangle = \sum_{\{i\}}\rho_i\ln(\rho_i/\rho_i^E)$, the 
\textit{mass} is equal to the stochastic Fisher information $-\partial^2_{\rho_i} \langle \Sigma^{ex} \rangle = 1/\rho_i$ 
defined with respect to $\rho_i$ \footnote{Note that we use the negative of the excess EP to 
ensure the positivity of the mass.}. Reorganization of 
$\Sigma_{drv}^{ex} = \sum_{\{i\}}(\partial_t\rho_i)^2/2\rho_i$ leads to 
$\Sigma_{drv}^{ex} = \sum_{\{i\}}\rho_i\left[\partial_t\ln(\rho_i/\rho_i^E)\right]^2$, the 
quadratic form of the excess driving EP written in terms of the driving of the excess affinity 
$A_i^{ex} = -\ln(\rho_i/\rho_i^E)$. Thus, the control of the state is equivalent to the 
control of the excess affinity. To avoid confusion with the state index, we introduce the 
shorthand notation $\vec{\rho}(\tau)$ and $\vec{\rho}(0)$ for the final and initial states of 
the system represented as component-wise vectors.

Using $\mathcal{L}_{drv}^{ex} = \sum_{\{i\}}(\partial_t\rho_i)^2/2\rho_i$ gives 
$\mathcal{G}(\rho_i) = 2\sqrt{\rho_i}$ and resolves 
$\Sigma_{drv}^{ex} = \int_0^\tau\mathcal{L}_{drv}^{ex} \, dt$. The corresponding slow--driving 
optimal excess EP and optimal driving protocol read:
\begin{subequations}\label{eq:slow_driving_excess_EP}
\begin{equation}\label{eq:slow_driving_optimal_excess_EP}
\begin{split}
    \Sigma_{qs}^{*ex} = \frac{2}{\tau}\left(\sqrt{\vec{\rho}(\tau)} - \sqrt{\vec{\rho}(0)}\right)^2,
\end{split}    
\end{equation}
\begin{equation}\label{eq:slow_driving_optimal_protocol_excess_EP}
\begin{split}
    \sqrt{\vec{\rho}(t)} & = \left(1 - \frac{t}{\tau}\right)\sqrt{\vec{\rho}(0)} + \frac{t}{\tau}\sqrt{\vec{\rho}(\tau)}. 
\end{split}    
\end{equation}
\end{subequations}
\Cref{eq:slow_driving_optimal_excess_EP} is the cEQ slow--driving formulation with a quadratic 
relation between the thermodynamic length and the excess EP \cite{Rao_1945,Qian_2002, Ito_2018, Loutchko_2022_prr_riemanian}. 
However, the corresponding optimal slow--driving protocol 
\cref{eq:slow_driving_optimal_protocol_excess_EP} is novel.

Furthermore, solving the GFTOC problem for the excess EP yields the finite--time optimal excess 
EP and finite--time optimal driving protocol:
\begin{subequations}\label{eq:finite_time_excess_EP}
\begin{equation}\label{eq:finite_time_optimal_excess_EP}
\begin{split}
    \Sigma_{\tau}^{*ex} &= \frac{2}{(2+\tau)}\left(\sqrt{\vec{\rho}(\tau)} - \sqrt{\vec{\rho}(0)}\right)^2,
\end{split}    
\end{equation}
\begin{equation}\label{eq:finite_time_optimal_protocol_excess_EP}
\begin{split}
    \sqrt{\vec{\rho}_\tau(t)} &= \left(\frac{1+\tau}{2+\tau} - \frac{t}{2+\tau}\right)\sqrt{\vec{\rho}(0)} + \left(\frac{1}{2+\tau} + \frac{t}{2+\tau}\right)\sqrt{\vec{\rho}(\tau)}.
\end{split}    
\end{equation}
\end{subequations}
\Cref{eq:finite_time_optimal_excess_EP} reveals that the `thermodynamic shock' lowers the 
driving cost of the excess EP due to \textit{kinks} in $\sqrt{\vec{\rho}_\tau(t)}$. Within 
an infinitesimal time at the initial and final driving times, the state of the system undergoes 
an instantaneous simultaneous jump. Importantly, 
\cref{eq:slow_driving_excess_EP,eq:finite_time_excess_EP} are constrained by the conservation 
law manifold and can therefore be further simplified or refined to express 
\cref{eq:slow_driving_excess_EP,eq:finite_time_excess_EP} as a function of the linearly 
independent state space, whose dimension is reduced by the total number of conservation laws. 

Since the choice of reference gauge $E$ is itself flexible and should be defined according to 
physical or experimental constraints, this formulation extends to any reference gauge; for 
example, $E = ss$ corresponds to any non-equilibrium steady state, with a focus on the 
optimization of the excess EP. 

{
$\Sigma_{qs}^{*ex}$ and $\Sigma_{\tau}^{*ex}$ are the `\textit{kinetic energetic}' finite--time dissipation 
corrections to the unavoidable quasi-static (`\textit{potential energetic}') excess EP change 
$\Delta_{\lambda_E} D_E^{KL} = D^{KL}_{E} ({\vec{\lambda}^{inl}}) - D^{KL}_{E}({\vec{\lambda}^{fnl}})$, which is usually assumed to be zero under the assumption that the initial and final state distributions satisfy the corresponding Boltzmann distributions. The case of $ D^{KL}_{E}({\vec{\lambda}^{inl}}) = 0$ (initially equilibrated state space Boltzmann distribution), and $ D^{KL}_{E} ({\vec{\lambda}^{fnl}}) \neq 0$ (cost penalty due to the final state--space distribution being unequal to the Boltzmann distribution dictated by the final control parameter) combined with the slow driving formulation [\cref{eq:slow_driving_excess_EP}], is predominantly studied in optimal transport theory and related machine learning applications. 
}
\subsection{Linear optimal control of housekeeping EPR}\label{sec:linear_optimal_control}
We aim to consider a novel formulation of optimal control of housekeeping EPR. Linear irreversible thermodynamics employs the most fundamental quadratic dissipation function 
$\mathcal{L}^* = kA_{\gamma}^2$ for EPR \cite{Onsager_1953, Onsager_1953_2}, which is also 
the regime associated with linear response theories in the non-equilibrium thermodynamics of NESS. This 
leads to $\mathcal{L}_{drv}^* = k\dot{A}_{\gamma}^2$ with \textit{mass} $\partial_{A_\gamma}^2 
\mathcal{L}^* = 2k$. Fixing $k = 1/2$ implies a unit mass. Solving $\dot{A}_\gamma = v_{qs}$ 
gives $\mathcal{G}^{lin}(A_\gamma) = A_\gamma$. Thus, the geodesic is a linear interpolation 
between $A_\gamma^i$ and $A_\gamma^f$ in $A_\gamma$ space, and the slow--driving optimal EP and 
corresponding optimal protocol read,
\begin{equation}\label{eq:linear_slow_optimal_process}
\begin{split}
    \Sigma_{qs}^* 
    &= \frac{1}{2\tau}\left(A_\gamma^f - A_\gamma^i\right)^2,
    \\
    A_\gamma(t) 
    & = \left(1 - \frac{t}{\tau}\right)A_\gamma^i + \frac{t}{\tau}A_\gamma^f,
\end{split}    
\end{equation}
respectively. The finite--time optimal driving EP and corresponding finite--time optimal protocol 
read,
\begin{equation}\label{eq:linear_finite_optimal_process}
\begin{split}
    \Sigma_\tau^* &= \frac{1}{2(2+\tau)}\left(A_\gamma^f - A_\gamma^i\right)^2,
    \\
    A_\gamma(t) 
    & =
    \left(\frac{1+\tau}{2+\tau} - \frac{t}{2+\tau}\right)A_\gamma^i + \left(\frac{1}{2+\tau} + \frac{t}{2+\tau}\right)A_\gamma^f.
\end{split}    
\end{equation}
The linear optimal control formulated here is a prototypical example of incorporating housekeeping EPR using Euclidean geometry with 
constant mass.

We aim to study the coarse--grained state--space representation of the transition--space description. Using the linear current-affinity relation $J_i (t) = 2 k A_i(t)$, valid in irreversible thermodynamics, and the state--space continuity equation $d_t \rho_i = J_i$ for $\rho_i$, this can be compactly expressed in vector notation as $d_t \vec{\rho} = \vec{J}$. Then, the state--space representation of \cref{eq:linear_slow_optimal_process} reads, 
\begin{equation}\label{eq:linear_state_slow_optimal_process}
\begin{split}
    \Sigma_{qs}^* 
    &= \frac{1}{2\tau} ||\vec{\rho}(\tau) - \vec{\rho}(0) ||^2,
    \\ 
    \vec{\rho}(t) 
    & = \left(1 - \frac{t}{\tau}\right) \vec{\rho}(0) + \frac{t}{\tau} \vec{\rho} (\tau),
\end{split}    
\end{equation}
The form of \cref{eq:linear_state_slow_optimal_process} resembles the speed limits in Refs.~\cite{Shiraishi_2018,Yoshimura_2021}; however, it is fundamentally different from them. The origin of \cref{eq:linear_state_slow_optimal_process} is geometric and is attributed to the \textit{kinetic--energetic} dissipation term; this notion is absent in Refs.~\cite{Shiraishi_2018,Yoshimura_2021}, which instead use the \textit{potential--energetic} dissipation contribution. Therefore, the speed limits in Refs.~\cite{Shiraishi_2018,Yoshimura_2021} belong to the equivalence class of the quadratic approximation of Eq.~(42) in Ref.~\cite{atm_2025_var_epr_derivation}, which addresses state--space speed limits derived under intrinsic dynamics with fixed control parameters. The finite-time state--space counterpart of \cref{eq:linear_finite_optimal_process} reads,
\begin{equation}\label{eq:linear_state_finite_optimal_process}
\begin{split}
    \Sigma_\tau^* & = \frac{1}{2(2+\tau)} || \vec{\rho} (\tau) - \vec{\rho} (0) ||^2,
    \\
    \vec{\rho}(t) 
    & =
    \left(\frac{1+\tau}{2+\tau} - \frac{t}{2+\tau}\right) \vec{\rho}(0) + \left(\frac{1}{2+\tau} + \frac{t}{2+\tau}\right) \vec{\rho} (\tau).
\end{split}    
\end{equation}

This example highlights that in the cEQ regime, where the housekeeping EPR 
substantially dominates the free energy and excess EP, a simple linear interpolation 
between control parameters yields a better optimal control procedure than that in 
\cref{sec:free_energy,sec:excess_epr}. Similar arguments apply if $J_i$ has a linear control-parameter dependence.
\subsection{Optimal control of housekeeping EPR}\label{sec:housekeeping_epr}
Using the housekeeping EPR $\langle \dot{\Sigma}^{hk} \rangle$ in \cref{eq:mean_EPR}, we formulate a novel and hitherto unresolved 
optimal control problem for $\langle \dot{\Sigma}_{hk} \rangle$, using external control of $F_\gamma$ from an 
initial to a final value $F_\gamma^i$ to $F_\gamma^f$. Physically, this implies that the 
control parameters $\{\lambda_E\}$ of $E$ are fixed and the state--space distribution is in 
steady state, leading to vanishing free energy driving and excess EPR contributions, respectively. In the following, the conjugate mobility $T_\gamma^\perp = 1$ is fixed, which ensures a unit \textit{mass} in the equilibrium limit $F_\gamma \to 0$. The 
driving Lagrangian for the housekeeping EPR reads $\mathcal{L}_{drv}^* = \frac{1}{2} 
\partial_{{F_\gamma}}^2\mathcal{L}_{hk}^*\left(\dot{F}_{{\gamma}}\right)^2$. The \textit{mass} for the 
driving reads $\partial_{{F_\gamma}}^2\mathcal{L}_{hk}^* = T_\gamma^{\perp}\cosh(F_\gamma/2) + 
T_\gamma^\perp F_\gamma\sinh(F_\gamma/2)/4$. It is rewritten as 
$\partial_{{F_\gamma}}^2\mathcal{L}_{hk}^* = \frac{1}{2}[T_\gamma(A_\gamma) + 
T_\gamma(A_\gamma^\dagger)] + \frac{1}{4} \langle \dot{\Sigma}_\gamma^{hk} \rangle$, with the total 
affinity $A_\gamma = F_\gamma + A_\gamma^{ex}$ and $A_\gamma^\dagger = F_\gamma - A_\gamma^{ex}$ 
being the affinity obtained by time-reversal of the boundary term. Using the orthogonal symmetry 
of the non-equilibrium fluctuations, $T_\gamma(A_\gamma) = T_\gamma(A_\gamma^\dagger)$, the 
\textit{mass} reduces to $\partial_{{F_\gamma}}^2\mathcal{L}_{hk}^* = T_\gamma + 
\langle \dot{\Sigma}_\gamma^{hk} \rangle /4$. Hence, for fEQ driving of $F_\gamma$, the \textit{mass} is proportional to 
the traffic $T_\gamma$ and the housekeeping EPR $\dot{\Sigma}_\gamma^{hk}$ due to the 
bidirectional transition $\gamma^\rightleftharpoons$, and increases exponentially as the system moves further from 
equilibrium due to the hyperbolic scalings of $T_\gamma$ and $\langle \dot{\Sigma}_\gamma \rangle$. This reveals 
that higher housekeeping EPR and non-equilibrium fluctuations generate higher driving resistance. 
Therefore, we use an exponential approximation of the hyperbolic functions, valid for fEQ 
systems, and use the previously computed closed-form analytical expression for the fEQ geodesic 
$\mathcal{G}^{fEQ}$ in \cref{eq:geodesic_expressions}. 

The slow--driving optimal control problem for the housekeeping EPR yields the optimal driving EP 
and corresponding optimal protocol,
\begin{equation}\label{eq:optimal_slow_driving_housekeeping_EP}
\begin{split}
    \Sigma_{qs}^{*hk} & = \frac{1}{2\tau}\left[ \mathcal{G}^{fEQ}(F_\gamma^f) - \mathcal{G}^{fEQ}(F_{\gamma}^i)\right ]^2,
    \\
    \mathcal{G}^{fEQ}(F_\gamma) & = \left(1 - \frac{t}{\tau}\right)\mathcal{G}^{fEQ}(F_\gamma^i) + \frac{t}{\tau}\mathcal{G}^{fEQ}(F_\gamma^f).
\end{split}    
\end{equation}
Similarly, the finite--time optimal control problem for the housekeeping EPR yields the 
finite--time optimal driving EP and corresponding finite--time optimal protocol,
\begin{equation}\label{eq:optimal_finite_time_housekeeping_EP}
\begin{split}
    \Sigma_{\tau}^{*hk} & = \frac{1}{2(2+\tau)}\left[ \mathcal{G}^{fEQ}(F_\gamma^f) - \mathcal{G}^{fEQ}(F_{\gamma}^i)\right]^2,
    \\
    \mathcal{G}_\tau^{fEQ}(F_\gamma) 
    & = \left(\frac{1+\tau}{2+\tau} - \frac{t}{2+\tau}\right)\mathcal{G}^{fEQ}(F_\gamma^{i}) 
    \\ & \hspace{1cm} + \left(\frac{1}{2+\tau} + \frac{t}{2+\tau}\right)\mathcal{G}^{fEQ}(F_\gamma^{f}).
\end{split}    
\end{equation}
A pictorial representation of the optimal control of $F_\gamma$ is shown in \cref{fig:2}, which 
pictorially summarizes the essence of this work.

{
To demonstrate the fEQ optimal control of the housekeeping EPR, we apply it to a prototypical 
analytically solvable three-state unicyclic (three-transition) graph, which forms the equivalence 
class of biological mechanisms and systems. Examples include ATP-to-ADP conversion in molecular 
motors \cite{Liepelt_2007}, enzymatic substrate-to-product catalysis \cite{Ninio_1975}, 
ion-channel gating \cite{Sigg_2014_ion_channels}, kinetic proofreading 
\cite{Hopfield_1974, Murugan_2012}, and gene expression regulation models. Following 
\cref{sec:setup}, $\{i\} = \{1, 2, 3\}$ and $\{\gamma^{\rightleftharpoons}\} = \{12, 23, 31\}$. 
We choose the control parameters $E = 0$ and $F_{\gamma} = F_{12} = F_{23} = F_{31} = F$. Due 
to the symmetry of this model, $\rho_i^{E} = \rho_i^{ss} = 1/3$ is independent of 
$F$ \footnote{Note that the state-density conservation constraint should be used for the CRN 
instead of the state--space probability conservation constraint 
($\sum_{\{i\}}\rho_i = 1$) used here for the MJP.}. The mean housekeeping EPR is 
$\langle\dot{\Sigma}^{hk}\rangle = 2(\rho_1 + \rho_2 + \rho_3)F\sinh(F/2) = 2F\sinh(F/2) 
= \mathcal{L}^*$, $F$ quantifies the directional asymmetry of the forward cyclic current $1\to2\to3\to 1$ and corresponding backward cyclic current $1\to3\to2\to1$. We consider the driving of the non-conservative affinity $F$ from an initial 
value $F^i = 2$ to a final value $F^f = 5$ (the same control parameter values as in 
\cref{fig:2}). As computed at the beginning of this section, 
$\partial_F^2\mathcal{L}_{hk}^* = 2\cosh(F/2) + \frac 1 2 F\sinh(F/2)$ and 
$\mathcal{L}_{drv}^{*} = \frac{1}{2}\partial_F^2\mathcal{L}_{hk}^*\dot{F}^2$. The optimal 
protocol and the corresponding finite--time driving EP are given by 
\cref{eq:optimal_finite_time_housekeeping_EP}, with a factor-of-2 difference in $\Sigma_\tau^*$ 
due to the definition of the mass.
}
\subsection{Multi--parameter optimal control of 
\\ affinity and mobility}\label{sec:optimal_control_of_mobility}
We demonstrate the applicability of multi-parameter optimal control by incorporating the 
symmetric part of the transition, namely the mobility $D_\alpha$, which was previously fixed. 
This implies $\{\lambda_\alpha^i\} = \{A_\alpha, D_\alpha\}$, $\forall\,A_\alpha \in 
\{A_\alpha\}$ with the corresponding $D_\alpha \in \{D_\alpha\}$, as evident from 
\cref{eq:mean_EPR} --- for \textit{Class~(1)}: $A_\alpha \in \{\lambda_E\}$ with $D_\alpha \in 
\{1\}$; for \textit{Class~(2)}: $A_\alpha \in \{-\ln(\rho_i/\rho_i^E)\}$ with $D_\alpha \in 
\{1\}$; and for \textit{Class~(3)}: $A_\alpha \in \{F_\gamma\}$ with $D_\alpha \in 
\{T_\gamma^\perp\}$ --- which are satisfied for the orthogonal decomposition of the affinities. 
If experimental constraints permit control over the set of microscopic affinities $\{A_\gamma\}$ 
and mobilities $\{D_\gamma\}$ instead of $\{A_\alpha, D_\alpha\}$, then the multi-parameter 
control for $\{\lambda_\gamma^i\} = \{A_\gamma, D_\gamma\}$ is defined analogously.
Using $\mathcal{L}^*$, we propose a novel finite--time optimal 
control problem for the simultaneous control of mobility $D_\alpha$ and affinity $A_\alpha$.
The exact metric tensor reads
\begin{widetext}
\begin{equation}\label{eq:metric_tensor_affinity_mobility}
\begin{split}
    g_{ij} 
    & =
    \begin{bmatrix}
        2D_\alpha\left(\cosh\!\left[\frac{A_{\alpha}}{2}\right] + \frac{A_\alpha}{2}\sinh\!\left[\frac{A_\alpha}{2}\right]\right) & \left(A_\alpha\cosh\!\left[\frac{A_\alpha}{2}\right] + 2\sinh\!\left[\frac{A_\alpha}{2}\right]\right)
        \\
        \left(A_\alpha\cosh\!\left[\frac{A_\alpha}{2}\right] + 2\sinh\!\left[\frac{A_\alpha}{2}\right]\right) & 0
    \end{bmatrix}.
\end{split}    
\end{equation}
\end{widetext}
Using \cref{eq:metric_tensor_affinity_mobility,eq:christoffel_symbol}, the exact Christoffel 
symbols read,
\begin{equation}\label{eq:christofel_symbol_}
\begin{split}
    \Gamma_{A_\alpha A_\alpha}^{A_\alpha} & 
    = \frac{\left[1 + \frac{A_\alpha}{4}\tanh\!\left(\frac{A_\alpha}{2}\right)\right]}{\left[A_\alpha + 2\tanh\!\left(\frac{A_\alpha}{2}\right)\right]}, 
    \\
    \Gamma_{D_\alpha A_\alpha}^{D_\alpha} & = \Gamma_{A_\alpha D_\alpha}^{D_\alpha} 
    = \frac{\left[1 + \frac{A_\alpha}{4}\tanh\!\left(\frac{A_\alpha}{2}\right)\right]}{\left[A_\alpha + 2\tanh\!\left(\frac{A_\alpha}{2}\right)\right]}, 
    \\
    \Gamma_{A_\alpha A_\alpha}^{D_\alpha} & = D_\alpha\frac{\left[-14 + A_\alpha^2 - 2\cosh(A_\alpha)\right]}{8\left[A_\alpha\cosh\!\left(\frac{A_\alpha}{2}\right) + 2\sinh\!\left(\frac{A_\alpha}{2}\right)\right]^2},
    \\
    \Gamma_{A_\alpha D_\alpha}^{A_\alpha} & = \Gamma_{D_\alpha A_\alpha}^{A_\alpha} = \Gamma_{D_\alpha D_\alpha}^{A_\alpha} = \Gamma_{D_\alpha D_\alpha}^{D_\alpha} = 0.
\end{split}    
\end{equation}
The exact geodesic equation \cref{eq:geodesic_equation_multi} for the optimal control of the 
transition affinity and mobility reduces to,
\begin{subequations}
\begin{equation}\label{eq:geodesic_affinity}
    \ddot{A}_\alpha + \Gamma_{A_\alpha A_\alpha}^{A_\alpha}\,\dot{A}_\alpha^2 = 0,
\end{equation}
\begin{equation}\label{eq:geodesic_mobility}
    \ddot{D}_\alpha + \Gamma_{A_\alpha A_\alpha}^{D_\alpha}\,\dot{A}_\alpha^2 + 2\Gamma_{A_\alpha D_\alpha}^{D_\alpha}\,\dot{A}_\alpha\dot{D}_\alpha = 0.
\end{equation}
\end{subequations}
\Cref{eq:geodesic_affinity} for the optimal driving of the affinity is identical to 
\cref{eq:geodesic_equation} and does not depend on $D_\alpha$. This is due to the linear dependence of $\mathcal{L}^*$ on $D_\alpha$. Therefore, the optimal driving 
dynamics of the mobility $D_\alpha$ given by \cref{eq:geodesic_mobility} are enslaved by the 
geodesic for $A_\alpha$. This implies that the optimal control problem for the transition 
affinity and mobility can be reduced to two steps. The first step is to solve the optimal 
control problem for $A_\alpha$, as formulated in \cref{sec:optimal_slow_driving,sec:finite_time_optimal_control}, 
and obtain the geodesic $\mathcal{G}(A_\alpha)$. The second step is to substitute 
$\mathcal{G}(A_\alpha)$ into \cref{eq:geodesic_mobility} and solve the ODE 
\cref{eq:geodesic_mobility} for the optimal driving of $D_\alpha$, obtaining the geodesic 
$\mathcal{G}(A_\alpha, D_\alpha)$. An exact analytical solution for the geodesic 
$\mathcal{G}(A_\alpha, D_\alpha)$ is not available due to the complexity of the coupled ODEs. 
However, numerical computation of $\mathcal{G}(A_\alpha, D_\alpha)$ is always feasible using 
\cref{alg:algorithm,alg:algorithm_action_minimization_optimal_protocols}. As evident from \cref{eq:metric_tensor_affinity_mobility}, this two--step reduction of the simultaneous optimal control of affinity and mobility is attributed to the linear dependence of $\mathcal{L}^*$ on $D_\alpha$; therefore, this simplification is shared by a more generic class of dissipation functions with the same linear-in-$D_\alpha$ structure. The finite--time 
optimal protocol $\mathcal{G}_\tau(A_\alpha, D_\alpha)$ then follows from 
$\mathcal{G}(A_\alpha, D_\alpha)$ and \cref{eq:optimal_geodescic_finite_time_multi_parameter}. 
\section{Conclusion and Outlook}\label{sec:conclusion_outlook}
We consider a finite--time optimal control problem for driving the control parameters of 
discrete-state systems from an initial to a final value, such that the driving EPR is minimised. 
Building upon the minimum action principle (MinAP), we propose the generalized finite--time 
optimal control (GFTOC) framework. MinAP allows for a variational formulation of the 
finite--time optimal control problem. To solve this problem, we develop a two-step solution. In 
the first step, assuming slow driving, we solve the slow--driving optimal control problem by 
exploiting the Riemannian geometry induced in the manifold of control parameters. This approach 
unifies and generalizes thermodynamic geometry and optimal transport theory for fEQ systems. The 
dissipation-minimizing optimal paths are then obtained via the geodesic, the minimum-distance 
path on the Riemannian manifold.

In the second step, we investigate the possibility of discontinuous endpoint jumps in the 
optimal protocol by relaxing the slow--driving assumption. We use the slow--driving geodesic to 
compute the finite--time geodesic; our analysis reveals an exact mapping between the slow--driving 
and finite--time optimal control problems. This strong and important result allows us to extend 
the framework of thermodynamic geometry, developed for slowly driven systems, to systems driven 
in any finite time. Importantly, due to MinAP, a thermodynamic cost is assigned to the endpoint 
jumps; this instantaneous dissipation cost at the endpoints is termed a `thermodynamic shock'. 
We analytically prove that it is a model-independent generic mechanism for far-from-equilibrium 
(fEQ) systems that reduces the driving dissipation cost. Within our framework, discontinuous 
endpoint jumps in the optimal protocol are a physical manifestation of a `thermodynamic shock', 
implying that the optimally driven dynamics in finite time are constrained by thermodynamics.

Our framework opens up a broad range of practical applications in biology, chemistry, population 
dynamics, and nanoscale/mesoscale devices, where stochastic thermodynamics has been an 
experimentally tested theoretical paradigm \cite{Ciliberto_2017}, opening the door for the 
experimental design, optimization, and control of such systems. The experimental verification of 
our framework awaits exploration \cite{Ma_2020}. The GFTOC framework is rather straightforwardly 
extended to quantum systems, as the development of optimal control and speed limits for quantum 
systems is attributed to the underlying geometric structure and not to the quantum nature 
\cite{Okuyama_2018, Shanahan_2018, Nicholson_2020}. The optimization of quantum systems also belongs to the equivalence class of excess EP (\cref{sec:excess_epr}). 
{The extensions of GFTOC to underdamped Langevin dynamics remains to be explored.} More broadly, the GFTOC methodology 
developed here in stochastic thermodynamics applies to other interdisciplinary domains that rely 
on Riemannian geometry as their mathematical foundation, including information geometry 
\cite{Rao_1945,Amari_2016_book}, in which the role of \textit{kinks} warrants further investigation.
\begin{acknowledgments}
ATM thanks Jin-Fu Chen for pointing out the experimental work Ref.~\cite{Ma_2020}, which 
motivated the theoretical formulation of the GFTOC framework.
\end{acknowledgments}

\textit{Note added}.\textemdash After releasing the first publicly available draft of this work, 
we noted that the GFTOC framework solves several major open problems identified in the 
`Perspectives and Conclusions' section of the recent review Ref.~\cite{Guery_Odelin_2023}, 
specifically those categorized as: `Geometric thermodynamic far-from-equilibrium shortcuts in 
swift state-to-state transformations'.

\textit
\appendix
\vskip 0.5cm
\subsection*{References}
\bibliography{reference}
\newpage
\end{document}